\documentclass[aps,prb,twocolumn,superscriptaddress,floatfix]{revtex4-1}
\usepackage[utf8]{inputenc}
\usepackage{graphicx}
\DeclareGraphicsExtensions{.png,.jpg,.pdf}
\usepackage{xcolor}
\usepackage{amsmath}
\usepackage{amssymb}

\usepackage{dcolumn}% Align table columns on decimal point
\usepackage{bm}% bold math
\usepackage{hyperref}
\hypersetup{colorlinks=true,linkcolor=blue,citecolor=blue}
\usepackage{float}

\newcommand{\ket}[1]{|#1\rangle}

\newcommand{\braket}[2]{\langle#1|#2\rangle}
\newcommand{\crea}[2]{#1^{\dagger}_{#2}}
\newcommand{\des}[2]{#1_{#2}}

\begin{document}

\title{Real space mapping of topological invariants using artificial neural networks}
\author{D. Carvalho}
\affiliation{Universidade de Aveiro, 3810-193 Aveiro, Portugal}
\author{N. A. Garc\'ia-Mart\'inez}
\affiliation{Quantalab, International Iberian Nanotechnology Laboratory (INL),
Av. Mestre Jos\'e Veiga, 4715-330 Braga, Portugal}
\author{J. L. Lado}
\affiliation{Quantalab, International Iberian Nanotechnology Laboratory (INL),
Av. Mestre Jos\'e Veiga, 4715-330 Braga, Portugal}
\affiliation{Institute for Theoretical Physics, ETH Zurich, 8093 Zurich, Switzerland}
\author{J. Fern\'andez-Rossier}
\thanks{On leave from Departamento de Física Aplicada, Universidad de Alicante, Spain}
\affiliation{Quantalab, International Iberian Nanotechnology Laboratory (INL),
Av. Mestre Jos\'e Veiga, 4715-330 Braga, Portugal}

\date{\today}

%%%%%%%%%%%%%%%%%%%%%%%%%%%%%%%%%%%%%%%%%%%%%%%%%%%%%%%%%%%%%%%%%%%%%%%%%%%%%%%%
\begin{abstract}
Topological invariants allow to characterize Hamiltonians, 
predicting the existence of topologically protected
in-gap modes.
Those invariants can be computed 
by tracing the evolution of
the occupied wavefunctions under twisted boundary conditions.
However, those procedures do not allow to calculate
a topological invariant by evaluating the system locally,
and thus require information about the wavefunctions in the
whole system.
Here we show that artificial neural networks can be trained to
identify the topological order by evaluating a local projection of the density
matrix.
We demonstrate this for two different models, a 1-D topological superconductor
and a 2-D quantum anomalous Hall state, both with spatially modulated
parameters.
Our neural network correctly identifies the different topological domains in
real space, predicting the location of in-gap states.
By combining a neural network with a calculation of the electronic states that
uses the Kernel Polynomial Method, we show that the local evaluation of the
invariant can be carried out by evaluating
a local quantity, in particular for systems without translational symmetry
consisting of tens of thousands of atoms. Our results show that supervised
learning is an efficient methodology to characterize the local topology of a
system.
\end{abstract}

\maketitle

%%%%%%%%%%%%%%%%%%%%%%%%%%%%%%%%%%%%%%%%%%%%%%%%%%%%%%%%%%%%%%%%%%%%%%%%%%%%%%%%
\section{Introduction}
The study of topological electronic phases is one of the central topics %themes
in modern Condensed Matter Physics. Depending on the symmetry class different
topological states exist, with the paradigmatic examples of
time reversal topological insulators,\cite{RevModPhys.82.3045}
topological superconductors,\cite{RevModPhys.83.1057}
topological crystal insulators,\cite{PhysRevLett.106.106802}
topological Kondo insulators\cite{PhysRevLett.104.106408} and
topological Mott insulators\cite{PhysRevLett.100.156401} among others.
The most fundamental quantity to characterize these states is the so called
topological invariant, whose value determines the topological class of the
system.
In particular, interfaces between systems with different topological invariants
show topologically protected excitations, resilient towards perturbations
respecting the symmetry class of the system.
Computationally, the calculation of the topological invariant usually requires
the explicit knowledge of the wavefunctions of the entire 
system.\cite{PhysRevB.83.235401,PhysRevB.95.075146,wu2017wanniertools}
In particular,
topological invariants can
be calculated as the winding number of the 
occupied wave functions 
under
twisted boundary
conditions.\cite{PhysRevB.83.235401,PhysRevB.95.075146,wu2017wanniertools}
In that way, these methods generically require computing the full
wavefunctions, that becomes a cumbersome task for
systems without translational symmetry consisting on thousands
of atoms.

In several situations of experimental relevance, translational symmetry is
broken and systems are able to show different phases in real space due to the
spatial modulation of the effective parameters.
This situation might lead to protected modes between different regions of the
system, dramatically changing the low energy properties of the whole material.
This is the natural scenario in van der Waals heterostructures, where Moire
patterns\cite{PhysRevB.90.075428,PhysRevB.96.085442,wang2016gaps} could coexist
with any topological state.\cite{sanchez2017helical,Young2014}
A more controlled situation is the proposals for topological superconductivity
involving nanowires, where the topological state is controlled \emph{locally} by
electric gates.\cite{Alicea2011,zhang2016ballistic}
Even though real space formulations for the topological invariant do
exist,\cite{PhysRevB.84.241106,PhysRevB.95.121114,loring2015k,mitchell2018amorphous,fulga2016aperiodic} their computation requires an
integration over the whole space. Thus,  there is not a simple
methodology to obtain a topological invariant  in inhomogeneus systems by
evaluating solely their local properties.

Application of Machine learning methods in Condensed Matter Physics
is a growing area.
A significant advantage of these techniques is that they are capable of finding the
important degrees of freedom of a dataset without needing a profound insight of
the treated problem.
The identification of phase transitions\cite{VanNieuwenburg2017,PhysRevX.7.031038,ohtsuki2016deep,PhysRevE.95.062122,broecker2017quantum,koch2017mutual}
and the study of the ground state and correlations in different quantum many
body
problems\cite{carleo2017solving,deng2016exact,PhysRevX.7.021021,PhysRevLett.118.216401,nagai2017self,PhysRevE.97.013306}
are just some of the problems that Machine Learning has helped tackle in the
past few years. Even some techniques have been used in combination with
\textit{ab initio} calculations allowing a broader and more accurate
understanding of
materials.\cite{PhysRevLett.108.058301,bartok2017machine,gao2016machine}
% Thus, our goal will be to borrow tools from machine learning to compute the
% topological invariant of a system based only on real space information.
Within the language of machine learning, the calculation of topological
invariants is understood as a simple classification algorithm,\cite{PhysRevLett.120.066401,yoshioka2017learning}
that could be
efficiently tackled with the so called artificial neural networks.
\cite{alexnet2012,Dede20107,Lecun1998,Goldberg2015,Bengio2003,pybrain2010jmlr}

In this manuscript we show that artificial neural networks (ANN) are capable of
characterizing the local topology of a system using as input a restricted amount
of real space information.
In particular, we show that a trained ANN identifies correctly the local
topological character in spatially varying Hamiltonians that create
topologically different regions in space.
Importantly, we show that this technique, used in conjunction with the kernel
polynomial method, allows to compute local topological invariants with
an algorithm whose computational cost scales just linearly with the size of the
system.

The rest of the paper is organized as follows. In section~\ref{sec:met} we
review the basics of artificial neural networks (Sec.~\ref{sec:NN}) and
summarize the use of the kernel polynomial method to efficiently compute density
matrices (Sec.~\ref{sec:KPM}).
In section~\ref{sec:Topo} we apply the combined ANN-KPM technique both to a
model Hamiltonian for a 1-D topological superconductor (Sec.~\ref{sec:1d}) and a
2-D anomalous Hall insulator (Sec.~\ref{sec:2d}). Finally, in
section~\ref{sec:Conc} we present our conclusions.

\section{Method}
\label{sec:met}
%%%%%%%%%%%%%%%%%%%%%%%%%%%%%%%%%%%%%%%%%%%%%%%%%%%%%%%%%%%%%%%%%%%%%%%%%%%%%%%%
\subsection{Artificial Neural Networks}
\label{sec:NN}
\begin{figure}[ht!]
\centering
\includegraphics[width=0.5\textwidth]{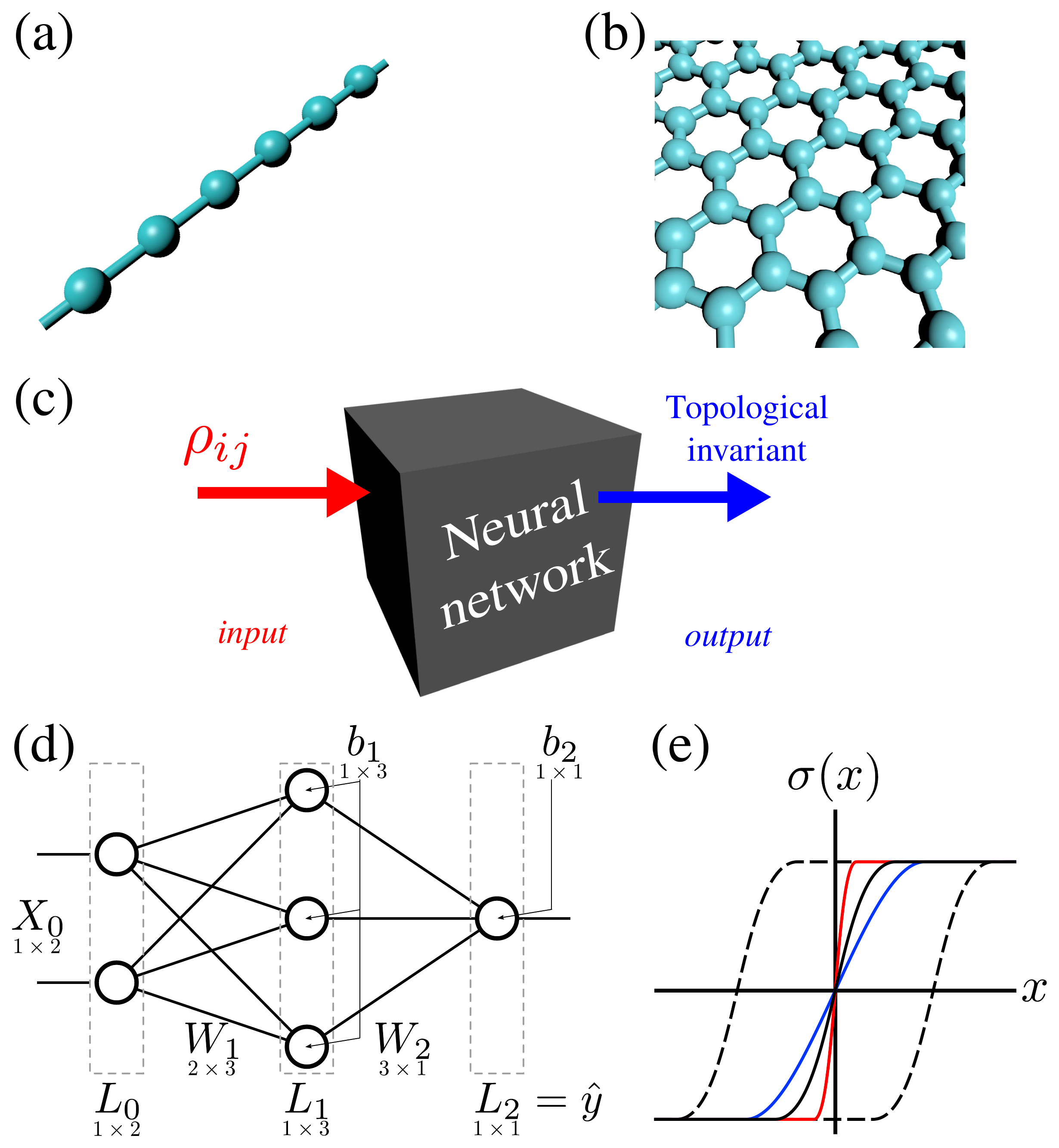}
\vspace{-5pt}
\caption{
Panels (a,b) show a cartoon of the two different geometries of the model Hamiltonians considered  below, 
%systems that we aim to topologically characterized.
a one dimensional topological superconductor  (a) and a two dimensional  quantum anomalous Hall insulator (b).
Panel (c) shows a schematic sketch of our procedure: a trained neural network
will take as input a local density matrix, and it will return the topological
invariant of the system.
Panel (d) shows a sketch for an artificial neural network as described in the text, whereas
in (e) we sketch the standard behavior of an activation function of a neuron,
$\sigma(x)$, for different weights (colors) and bias (dashed lines).
}
\label{ANN}
\label{fig1}
\end{figure}

%% In this paragraph we should explain:
%  - NN are a only subset of ML
%  - Supervised learning algorithms
%     `- Model to correlate inputs-outputs
Machine Learning (ML) is a broad field that includes many different approaches,
goals and methods.\cite{Solomonoff1957}
The defining property of ML algorithms is that they allow computers to perform
specific tasks without being explicitly programmed for each one of
them.\cite{Samuel1959} Within the vast variety of ML algorithms, we will focus
on supervised learning algorithms, which require a training dataset to fit the
parameters in the model. One of the most common models of supervised learning are
\emph{Artificial Neural Networks} (ANN) which have been proven very useful to
model patterns and correlations of complex problems that cannot be modeled
analytically such as image or sound
recognition,\cite{alexnet2012,Dede20107,Lecun1998} and even natural language
processing.\cite{Goldberg2015,Bengio2003}
In our case, we aim to use an artificial neural network to characterize
locally the topological state of a one (Fig.~\ref{fig1}~(a)) or two
(Fig.~\ref{fig1}~(b)) dimensional system. The objective of the procedure
is to have a neural network that, given local information about the system,
returns the topological invariant as sketched in Fig.~\ref{fig1}~(c).
The local information that will be provided is a local block of the density
matrix of the system, as we will discuss later.

%~~~~~~~~~~~~~~~~~~~~~~~~~~ FIGURE ~~~~~~~~~~~~~~~~~~~~~~~~~%
%~~~~~~~~~~~~~~~~~~~~~~~~~~~~~~~~~~~~~~~~~~~~~~~~~~~~~~~~~~~%

ANN are loosely based on parts of the brain, consisting of neurons, modeled as
perceptrons\cite{Rosenblatt1958}, and synapses as shown in Fig.~\ref{ANN}~(a).
%\buemark{Every neuron in an ANN has a number of inputs and outputs in the form of real numbers}
The neurons in an ANN do not attempt to model the actual structure or behavior
of the biological cells\cite{Hodgkin1952}. Instead, they mimic one of their main
features, the activation function.
This activation function, $\sigma$, sketched in Fig.~\ref{ANN}~(b) provides the
output of each neuron based on the received inputs and an external parameter
(bias). For computational convenience, $\sigma$ should be any smooth and
differentiable function defined over $\mathbb{R}$ but with its range restricted
to a closed interval, namely $\sigma\in[-1,1]$, as depicted in Fig.~\ref{ANN}~(e).
Usually, these functions are either the $\tanh$ or the sigmoid function, but
others might be used without loss of generality or functionality since these
models are only weakly sensitive to these details.\cite{Hopfield1982}

The inputs $X$ entering each neuron are weighted by the synapses $W$ and shifted
by the bias $b$. The synapses' weights are parameters to be tuned and they can be
arranged as rectangular matrices, $W^{\alpha}$, so the output
$L$ of the layer $\alpha$ can be obtained simply as:
$L_\alpha=\sigma(X_\alpha\cdot W_\alpha + b_\alpha)$,
where $X_\alpha$ is the input of the layer $\alpha$ (note that for the hidden
layers $X_\alpha=L_{\alpha-1}$).
As a formative example the outputs of every layer of the toy model sketched in
Fig.~\ref{ANN}~(d) can be calculated as follows:
\begin{equation}
  \begin{split}
    L_0 &= \sigma(X_0) \quad\text{or just}\quad L_0 = X_0 \\
    L_1 &= \sigma\left(L_0 \cdot W_1 + b_1\right) \\
    L_2 &= \sigma\left(L_1\cdot W_2 + b_2\right)= \hat{y}
  \end{split}
\label{FF}
\end{equation}
where $X_0$ is the input fed to the ANN.
The matrices $W_\alpha$ and the arrays $b_\alpha$ are the parameters to be
fitted during the training process in order to modify the activation functions
of each of the neurons in the ways showed in Fig.~\ref{ANN}~(e). Note that the
number of parameters in ANN models grows very quickly with the size (number of
neurons per layer) and depth (number of layers) of the network.\\

%In our case, we will employ ANN to implement a supervised
%learning algorithm. This kind of algorithms consist in three phases.
Artificial neural network, as every supervised learning algorithm, consist in
three phases.
First, the architecture of the model (i.e. the number of layers and neurons per
layer) is decided depending on the complexity of the problem addressed.
Second, the model is trained. In this process, several input-output pairs are
provided to the model whose parameters are fitted to mimic the correlations
present in the user-provided data.
Finally, when the training is completed, the model can be used to evaluate new
(unseen) input data.

Supervised learning algorithms require a training dataset to optimize the
parameters of the models. The training is performed by minimizing a cost
function, $\mathcal{E}$, usually proportional to the squared difference between
the expected output, $y$, and the actual output of the network, $\hat{y}$.
\begin{equation}
  \mathcal{E}=\tfrac{1}{2}(y-\hat{y})^2
\end{equation}
Notice that $y$ is a constant defined by the (user-provided) training dataset
while $\hat{y}$ depends on all the parameters of the network (weights and bias).
The minimization of $\mathcal{E}$ is, then, performed by iteratively modifying
the values of all the weights and bias in the network until the desired output
is obtained.
This is a computationally complex and expensive process since the number of
parameters can range from a few tens to millions. In fact, it was not until 1986
that an efficient method was developed for such a purpose.\cite{Rumelhart1986}
We use the gradient descent with the back-propagation algorithm to train the
ANN, which is the most common approach nowadays.
We used the open source library PyBrain,\cite{pybrain2010jmlr} to create, train and evaluate the ANN.

%~~~~~~~~~~~~~~~~~~~~~~~~~~ FIGURE ~~~~~~~~~~~~~~~~~~~~~~~~~%
\begin{figure}[t!]
\centering
\includegraphics[width=\columnwidth]{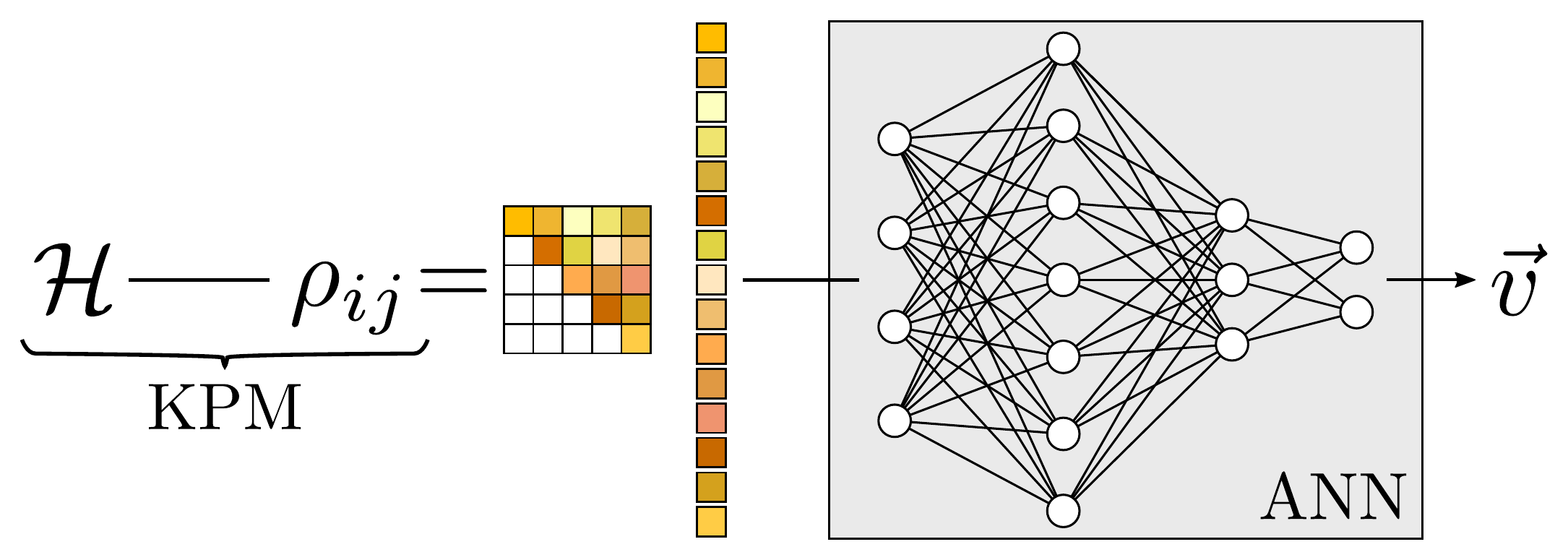}
\vspace{-5pt}
\caption{
Sketch of the process to evaluate the topological character of a local region
of space. The density matrix corresponding to a certain area in space is
calculated using the KPM, after removing the redundant elements the matrix is
rearranged in a 1D array that is used as input for neural network that will
provide the corresponding topological invariant as output.
}
\label{fig_method}
\end{figure}
%~~~~~~~~~~~~~~~~~~~~~~~~~~~~~~~~~~~~~~~~~~~~~~~~~~~~~~~~~~~%

\subsection{Correlation functions with the Kernel Polynomial method}
\label{sec:KPM}
In this section we review how real space correlation functions can be
efficiently calculated using the Kernel polynomial method
(KPM).\cite{RevModPhys.78.275}
We will focus the discussion in the case of a normal electronic system,
since the case of a superconductor can be treated in an analogous way.
The main task that we have to perform is to obtain the density matrix, evaluated
in a restricted area of real space, of a certain (very large) Hamiltonian. In
terms of the eigenfunctions $|\Psi_k\rangle$ of the Hamiltonian $H$,
the elements of the density
matrix can be written as
\begin{equation}
\rho_{ij} = \int_{-\infty}^{E_F} \langle i | \Psi_k \rangle \langle \Psi_k | j\rangle\delta(E_k -\omega) d\omega
\label{dens}
\end{equation}
where $|i\rangle$ and $|j\rangle$ are the elements of the basis for the
Hamiltonian $H$ and $E_F$ is the Fermi energy.
The diagonal elements of the matrix, $\rho_{ii}$, are the integrated local
density of
states.
In the gapped state, the off-diagonal elements are expected to decay
exponentially with distance. So, when the Fermi energy $E_F$ lies in the gap,
the density matrix is properly described by restricting the calculation to a set
of neighboring sites.
Generically, calculating the previous matrix requires diagonalizing the full
Hamiltonian to obtain the occupied wavefunctions, a task that scales with
$\mathcal{O}(N^3)$, with $N$ the system size of the system.
The Kernel Polynomial Method allows the computation of $\rho_{ij}$, for a
restricted set of neighboring sites, with a computational cost that scales only
as $\mathcal{O}(N)$.

The KPM allows to compute the quantity
\begin{equation}
g_{ij}(\omega) = \sum_k \langle i | \Psi_k \rangle \langle \Psi_k | j\rangle\delta(E_k -\omega)
\end{equation}
which can easily be integrated to obtain the density matrix~\eqref{dens}. The
central idea is that $g_{ij}$ can be expressed in matrix form as
\begin{equation}
g_{ij}(\omega) =
\langle i | \delta (H-\omega) | j \rangle
\label{eq5}
\end{equation}

The KPM consists on expanding equation~\eqref{eq5} in terms of Chebyshev
polynomials $T_n(\omega)$. To do so, the Hamiltonian is first rescaled so that
all the eigenenergies lie in the interval $\mathcal{E}_k \in (-1,1)$. The
rescaled Hamiltonian is denoted as $\mathcal{H}$. The corresponding spectral
function is calculated as
\begin{equation}
%g_ij{\bar\omega} typo?
g_{ij}({\omega}) = \frac{1}{\pi \sqrt{1-\omega^2}}
\left (\bar \mu_n + 2 \sum^N_{n=1} \tilde \mu_n T_n (\bar \omega)
\right )
\label{KPM}
\end{equation}
The coefficients $\tilde \mu_n$ determine the expansion of a certain element
$g_{ij}$, and are calculated as

\begin{equation}
\tilde \mu_n = g^N_n \mu_n
\end{equation}
where $\mu_n$ are the coefficients calculated from the Hamiltonian
$\mathcal{H}$ and $g_n^N$ denotes the Jackson Kernel that improves the
convergence of the series\cite{RevModPhys.78.275}

\begin{equation}
g_n^N =
\frac{(N-n-1)\cos \frac{\pi n}{N+1} + \sin \frac{\pi n}{N+1}
\cot \frac{\pi }{N+1}
}
{N+1}
\end{equation}

Given two sites $i$ and $j$, we define two vectors located in those sites
$v_i$ and $v_j$.
The coefficients $\mu_n$ would be calculated as a conventional functional
expansion

\begin{equation}
\mu_n = \langle v_i | \int_{-1}^{1}\delta (\mathcal{H}-\omega)T_n(\mathcal{H}) d \omega | v_j \rangle
\end{equation}
which in the diagonal basis reads
\begin{equation}
\mu_n = \int_{-1}^{1}\langle v_i | \Psi_k \rangle \delta (\mathcal{E}_k-\omega) \langle \Psi_k | v_j \rangle T_n(\omega) d \omega
\end{equation}
Performing the integration over $\omega$ we get
\begin{equation}
\mu_n = \langle v_i | \Psi_k \rangle T_n(\mathcal{E}_k)  \langle \Psi_k | v_j \rangle =
\langle v_i | T_n(\mathcal{H}) | v_j \rangle
\end{equation}
Therefore, the coefficients $\mu_n$ can be calculated as the overlap of two
vectors

\begin{equation}
\mu_n =
\braket{v_j}{\alpha_n}%\langle \alpha_0 | \alpha_n \rangle
\end{equation}
where $\alpha_n$ is calculated with
the recursion relations associated to the Chebyshev polynomials
\begin{equation}
\begin{aligned}
\ket{\alpha_0} = \ket{v_i}  \\  %|\alpha_0 \rangle = | v_i \rangle \\
|\alpha_1 \rangle = \mathcal{H} | \alpha_0 \rangle \\
|\alpha_{n+1} \rangle = 2\mathcal{H} | \alpha_n \rangle-
| \alpha_{n-1} \rangle
\end{aligned}
\end{equation}
This procedure thus involves matrix vector products to calculate the
coefficients. For a sparse matrix, as it is the case of a tight binding
Hamiltonian, the number of non-zero elements scales linearly with the system
size, so the computational cost of calculating the density matrix for a fixed
number of sites also scales linearly.
This method allows to compute $g_{ij}$ at every energy simultaneously, so
that $\rho_{ij}$ can be calculated by integration up to the Fermi energy.

For small systems, the density matrix can be calculated also by exact
diagonalization of the full Hamiltonian. In principle, that procedure
allows to calculate the
correlation function of relatively large one dimensional systems. However, for a
two dimensional system, the dimension of the matrix will be too large in
general. It is in that situation when the kernel polynomial method is specially
suitable.

\section{Topological invariants with supervised learning}
\label{sec:Topo}
We now describe the procedure to characterize the local topological character of
a system using ANN.
We choose as input the elements of the density matrix that involve one site and
its closest neighbors.
This procedure allows to naturally treat systems without translation
symmetry and with disorder, as the calculation of the density matrices is not more
computationally expensive in those situations using the previous procedure.
% Noel: This sentence was a bit weird for me
The process to calculate the density matrix was discussed in
Sec.~\ref{sec:KPM} but in order to use the density matrix as input for our ANN
some processing is required.
The density matrix is, in general, a complex Hermitian matrix, so we will
remove the redundant elements (namely the lower triangle) and arrange the
remaining elements in a 1D array, concatenating the real and imaginary parts.
Furthermore, we included the eigenvalues of the density matrix as part of the
input. Strictly, the inclusion of the eigenvalues is a redundant operation that
could be avoided by increasing the size and/or depth of the ANN, yet we found
that it helped the optimization of the model with a negligible computational
overhead.
The output of the NN will be the topological invariant of the system.
The calculation of the corresponding output is done by constructing a
translational invariant Hamiltonian in which the corresponding topological
invariant is well defined and can be calculated in a standard way.
Finally, since we are using the ANN as a classifier, it is convenient to encode
the possible outputs as linearly independent vectors, $\vec{v}$, rather
than use a single scalar. The use of vectors allows the discrimination
between wrong answers and false positives.
This whole architecture is sketched in Fig. \ref{fig_method}.

In order to train of the ANN we generate a large number of realizations of a
family of Hamiltonians, exploring their parameter space. For a given choice of
parameters, we compute the topological invariant of the corresponding pristine
case and its local density matrix.
These procedure allows us to generate a set of inputs and outputs, which are
used to train the NN.
Once the ANN is trained, the model is ready to be evaluated with new data that
the network has never tried to test the accuracy of the network.
The last step is to create a new Hamiltonian with spatially dependent
parameters, and evaluate the NN with the local density matrix corresponding to a
neighborhood of every lattice site.
In this way, we have a procedure to locally evaluate the topological
invariant of a systems lacking translational symmetry.

\subsection{One dimensional topological superconductor}
\label{sec:1d}
%~~~~~~~~~~~~~~~~~~~~~~~~~~ FIGURE ~~~~~~~~~~~~~~~~~~~~~~~~~%
\begin{figure}[t!]
\centering
\includegraphics[width=0.5\textwidth]{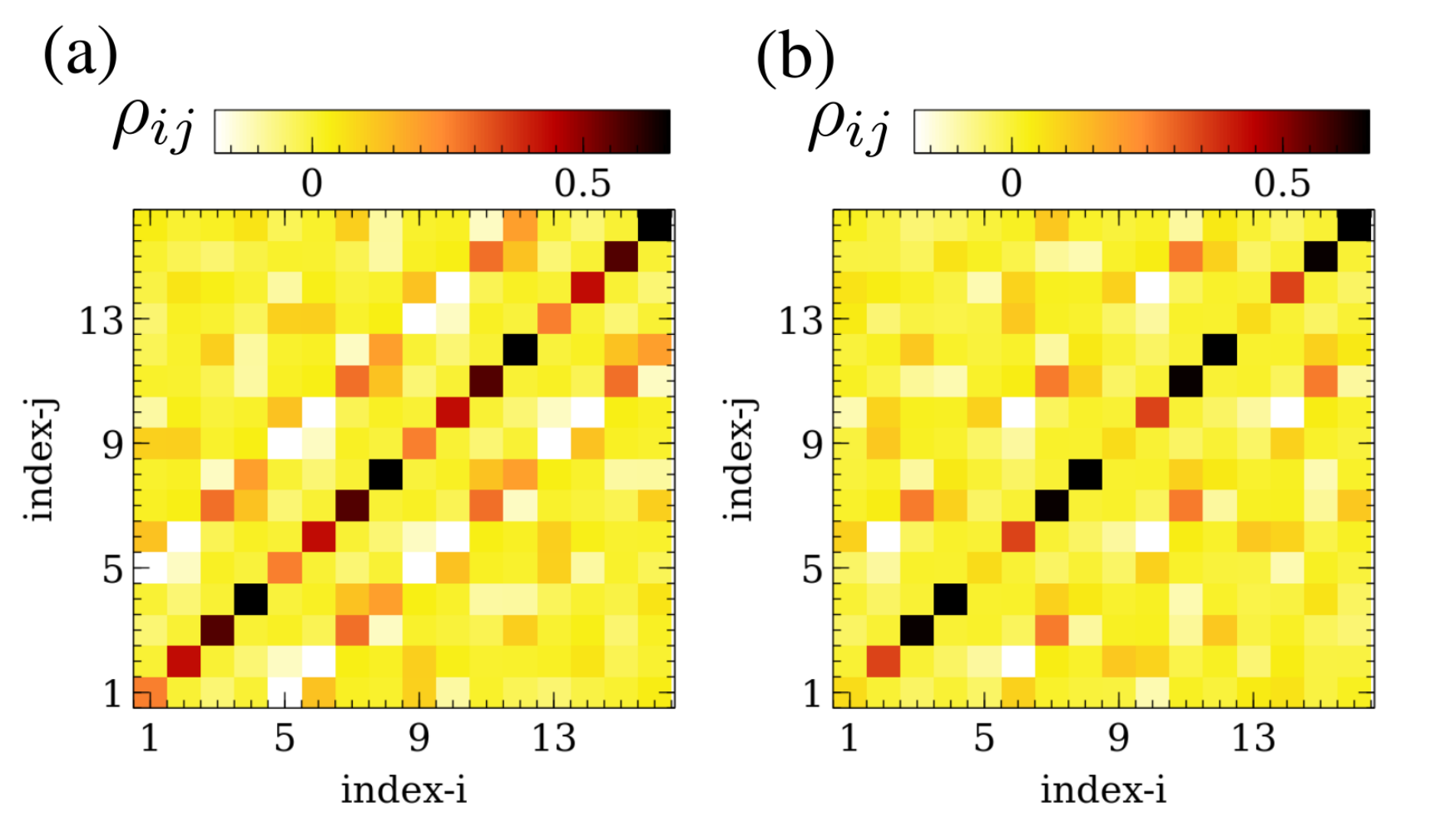}
\vspace{-5pt}
\caption{
%Matrix
Image representation of the density matrix for two particular different states
for the superconducting 1D system, trivial (a) and topological (b). In terms of
the matrices shown, the task of the neural network can be understood as an image
recognition algorithm capable of distinguishing an input (a) from (b), for
different parameters of the Hamiltonian chosen randomly. The different indexes
in x and y axis run over spin and electron/hole sectors in the closest sites.
}
\label{fig3}
\end{figure}
%~~~~~~~~~~~~~~~~~~~~~~~~~~~~~~~~~~~~~~~~~~~~~~~~~~~~~~~~~~~%

In the following, we will consider a lattice model Hamiltonian for a one
dimensional electron gas that is able to host both trivial and topological
superconducting states. The corresponding topological invariant is a $Z_2$
number that can be calculated as a Berry phase.\cite{PhysRevB.88.075419}
Such effective one dimensional system, in particular the superconducting
topological phase, is realized in semiconducting nanowires deposited on top of a
s-wave
superconductor.\cite{PhysRevLett.105.077001,PhysRevLett.105.177002,mourik2012signatures,PhysRevB.84.144522,PhysRevLett.106.127001,lutchyn2017realizing,aguado2017majorana}
The model describes electrons in a 1D chain, in the presence of Zeeman
field, Rashba spin-orbit coupling, superconducting proximity effect
and a sublattice imbalance term.
Thus, the model has six different parameters: a spin-conserving hopping $t$,
chemical potential $\mu$, Rashba spin orbit $t_R$, external Zeeman field $B_z$,
on-site pairing term $\Delta$ and a trivial mass $m$.
Moreover, we also include the possibility of having finite Anderson disorder
$W_i$,
so that the full Hamiltonian reads

\begin{equation}
\begin{split}
  \mathcal{H} =&-t\sum_{\langle ij\rangle_\alpha}
                \crea{c}{i\alpha}\des{c}{j\alpha}
                 + i t_R\sum_{\langle ij\rangle_{\alpha\beta}}
      \hat{e}_z\cdot(\vec{\sigma}_{\alpha\beta}\times\vec{d}_{ij})
                                           \crea{c}{i\alpha}\des{c}{j\beta}\\
      &+B_z\sum_{i\alpha}\crea{c}{i\alpha}\sigma_z\des{c}{i\alpha}
  +\Delta \sum_i [c_{i\uparrow} c_{i\downarrow} + c^\dagger_{i\downarrow} c^\dagger_{i\uparrow}]\\
  &+ \mu\sum_{i,\alpha}\crea{c}{i\alpha}\des{c}{i\alpha}
  +m\sum_{i}\tau_i\crea{c}{i}\des{c}{i}
  +\sum_{i,\alpha} W_i \crea{c}{i\alpha}\des{c}{i\alpha}
  \label{hamil1d}
\end{split}
\end{equation}

The previous Hamiltonian can have topological and trivial phases.
In a nutshell, a topological phase may arise when the Zeeman term $B_z$ is such
that the chemical potential $\mu$ crosses only one of the spin channels, so that
a small pairing $\Delta$ and Rashba field $t_R$ gives rise to a spinless p-wave
superconductor.\cite{PhysRevLett.106.127001}
In the absence of both Zeeman and Rashba couplings, the induced superconducting
gap is trivial.

The Hamiltonian \eqref{hamil1d} is solved in the Nambu representation by defining
a spinor wavefunction as
$
%\begin{equation}
\Psi^\dagger =
\begin{pmatrix}
c^\dagger_{\uparrow}, &
c^\dagger_{\downarrow} ,&
c_{\downarrow}, &
-c_{\uparrow}
\end{pmatrix}
%\end{equation}
$
which gives rise to a Bogoliuvov-de-Gennes Hamiltonian
$\mathcal{H} = \frac{1}{2}\Psi^\dagger H \Psi$.
The matrix $H$ is used to calculate the correlation functions
$\langle c_{i,s} c_{j,s'} \rangle$
and
$\langle c^\dagger_{i,s} c_{j,s'} \rangle$,
as introduced in section~\ref{sec:KPM}, by integrating the different
$g_{ij}(\omega)$ from $\omega=-\infty$ up to $\omega=0$.
In Fig.~\ref{fig3} we show an example of two different input data from the
training dataset, for a topological (a) and a trivial (b) state computed for an
open chain with $N=400$ sites using the KPM.
It is evident that simple inspection is not enough to distinguish between the two
of them. 
%Interestingly, the lack of an analytic description for the
%topological invariant based solely in the local information makes the
%recognition of the different phases a non-trivial task, that our ANN is able to
%handle.

\begin{figure}[t!]
\centering
\includegraphics[width=0.45\textwidth]{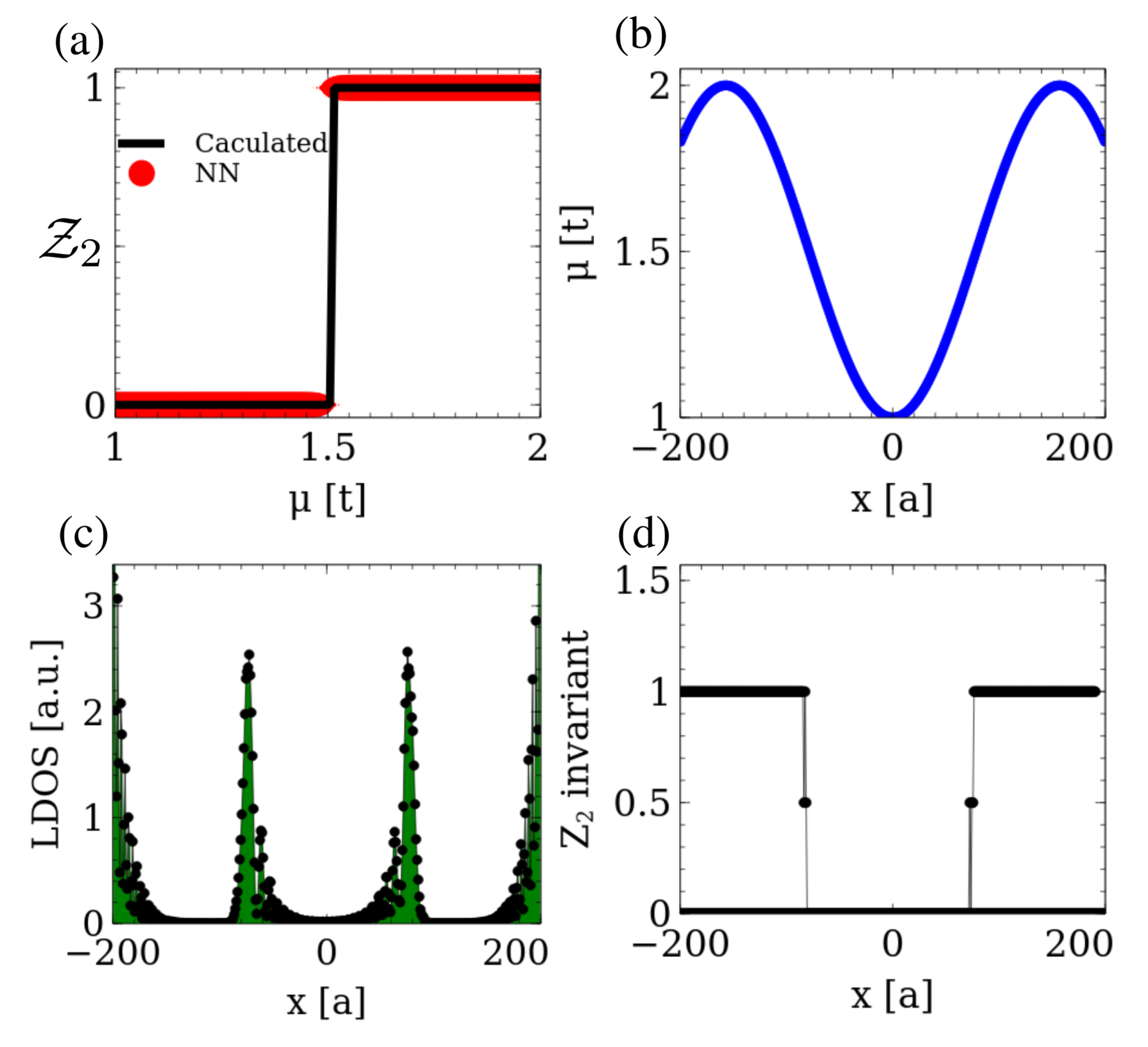}
\vspace{-5pt}
\caption{
(a) Comparison of the topological invariant computed exactly with the one
predicted by the trained neural network in a pristine system,
showing that the ANN perfectly captures the phase transitions
in the homogeneus system. Afterwards, we
create a inhomogeneus system with modulated chemical potential as shown in
panel (b).
Such modulation creates trivial and topological zones, with Majorana modes
pinned at the transition points (c).
The neural network is then evaluated in every point of the space, yielding a
site-dependent topological invariant shown in (d). The topological
transitions shown in (d) mark the existence of zero Majorana modes obtained
in (c).
The parameters used are $\lambda_R=-0.3t$
%($Z_{ee,x}=Z_{ee,y}=0$,
$B_{z}=0.5t$, $m=0$, and $\Delta =0.1t$.
}
\label{fig4}
\end{figure}

In order to generate the training dataset we considered different Hamiltonians
for a bipartite chain with 400 sites by varying the different values for the
off-plane Zeeman field $B_z$, Rashba $\lambda_R$, chemical potential $\mu$, 
superconducting pairing $\Delta$ and sublattice imbalance $m$.
In order to prove the robustness of our procedure, we also
switch on the Anderson on-site
disorder ($W\in (0.0,0.4t)$), with
a magnitude comparable to the other energy scales.
For the training dataset we generated 1000 different Hamiltonians with
parameters randomly chosen in the following ranges:
$t_R\in[-0.3t,0]$, $B_z\in[0.2t,0.8t]$, $\mu\in[t,2t]$, $\Delta\in[0.1t,0.3t]$,
$m\in[-0.2t,0.2t]$, 
yielding a five dimensional phase space. Using the generated Hamiltonians we
calculate the density matrix of the central atom in the nanowire, $\rho_{ij}$,
and its three closest neighbors.
Since the Hamiltonian in eq.~\eqref{hamil1d} only involves two Pauli matrices for
a linear chain, the Hamiltonian in real space can be chosen to be purely real,
so that its density matrix will be also real.
For each example the $Z_2$ topological invariant is calculated for the pristine
system ($W_i=0$)
defined by that particular set of parameters, which is used as expected
output. Since this topological invariant only has two possibilities, we encode
the $Z_2$ invariant as a two dimensional vector $v$, so that the topological
case corresponds to $v=(1,0)$ and the trivial case to $v=(0,1)$.
With this methodology a single element of the training dataset has a
152-dimensional input and a 2-dimensional output.
We took two hidden layers with 101 and 21 neurons.
After training, a validation set with 200 new samples is generated to test the
accuracy of the ANN yielding an accuracy of $\sim97\%$.
In order to gain some insight on the ANN capabilities, we run a simple test by
freezing all the parameters in the Hamiltonian~\eqref{hamil1d} but the chemical
potential and comparing the actual $Z_2$ with the output provided by the ANN.
In Fig~\ref{fig4}~(a) we see that even for unseen data the NN is able to provide
the correct topological invariant.

Once the network is trained, it is ready to be used in the case of an inhomogeneus
system.
We now generate a one dimensional system following equation~\eqref{hamil1d} with
spatially varying couplings.
In particular, we modulate the chemical potential along the chain as shown in
Fig.~\ref{fig4}~(b). Such modulation is feasible by means of local gates in the
experimental realizations involving semiconducting
nanowires.\cite{zhang2016ballistic}
With such modulation, we observe the emergence of zero energy modes in the local
density of states~\ref{fig4}~(c), which are expected to be a signature of a
boundary between a trivial and topological phase.
The evaluation of the topological invariant on every atomic position of the
chain can be carried out by feeding the local density matrix to the trained
neural network.
Our network shows that the different regions of the space have different
topological invariants as shown in Fig.~\ref{fig4}~(d).
It is observed that the points of space where the topological invariant changes
in Fig.~\ref{fig4}~(d) correspond to the location of the zero energy Majorana
modes, as seen in Fig.~\ref{fig4}~(c), validating the performance of our neural network.

%~~~~~~~~~~~~~~~~~~~~~~~~~~ FIGURE ~~~~~~~~~~~~~~~~~~~~~~~~~%
%~~~~~~~~~~~~~~~~~~~~~~~~~~~~~~~~~~~~~~~~~~~~~~~~~~~~~~~~~~~%

The success of the neural network in describing the topological order of the
different phases implies that, locally, the density matrix carries
enough information to distinguish between the two cases. In particular, the
elements of $\rho_{ij}$ involving $\langle c_{i,s} c_{j,s'}\rangle$ encode
information about the induced superconducting order parameter, both in the $s$
and $p$-wave channels, which physically is expected to determine the topological
phase.
If, in comparison, only the diagonal part of the density matrix was used
as input for the neural network, it would not be possible to distinguish between
trivial and topological states.
This is easily understood taking into account that the diagonal part of
$\rho_{ij}$ accounts for the total occupation numbers and two topologically
inequivalent band-structures can have arbitrarily similar density of states.

\subsection{Two dimensional Chern insulator}
\label{sec:2d}
In this section we will use an analogous methodology to study a topological two
dimensional state. In particular, we consider a model Hamiltonian for
electrons moving in a honeycomb lattice with Rashba spin orbit coupling $t_R$,
off-plane exchange $B_z$, that is known to result in a two dimensional Quantum
Anomalous Hall state (QAH)\cite{Qiao2010}:
\begin{equation}
\begin{array}{c@{}l}
  H = -t\sum_{\langle ij\rangle_\alpha}\crea{c}{i\alpha}\des{c}{j\alpha}
    + i t_R\sum_{\langle ij\rangle_{\alpha\beta}}
    \hat{e}_z\cdot(\vec{\sigma}_{\alpha\beta}\times\vec{d}_{ij})
                                         \crea{c}{i\alpha}\des{c}{j\beta}\\
    +B_z\sum_{i\alpha}\crea{c}{i\alpha}\sigma_z\des{c}{i\alpha}
    % +\lambda_{m}  % for consistence with the other equation
    +m\sum_{i,\alpha}\tau_i\crea{c}{i\alpha}\des{c}{i\alpha}  \\
  +\sum_{i,\alpha} W_i \crea{c}{i\alpha}\des{c}{i\alpha}
  \end{array}
\label{hamil:2d}
\end{equation}
where $t_R$ is the Rashba coupling, $\vec{\sigma}$ are the spin Pauli matrices,
$B_z$ is the external Zeeman field and $\tau_i=\pm1$ is the sublattice operator.
The first term is the usual tight-binding hopping term, the second one describes
the Rashba interaction~\cite{Qiao2010,Min2006} and the third term is the
so-called exchange or Zeeman term which couples to the spin degree of freedom.
The fourth term is a trivial mass term that assigns an opposite on-site
energy for the atoms in each of the sublattices, that we introduce in order
to have a trivial insulator phase in the model.
Finally, the last term is an Anderson disorder term that we introduce
to prove the robustness of the procedure. For $m=0$, and $B_z\neq0$ and
$t_R\neq 0$, the model has a topological gap with a  with Chern number
$\mathcal{C}=\pm 2$. For $m\neq 0$ and $B_z=0$. the model has a trivial
($\mathcal{C}=0$) gap.

Each of these Hamiltonian terms can effectively describe different experimental
situations. The sublattice imbalance could arise for a graphene monolayer
deposited on boron nitride in a commensurate
fashion.\cite{wang2016gaps,PhysRevB.88.035448}
The Rashba and exchange fields naturally arise for a graphene monolayer
deposited over a ferromagnetic insulator, such as
YIG,\cite{tang2017approaching,PhysRevLett.114.016603}
EuO\cite{PhysRevB.95.075418} or
CrI$_3$.\cite{huang2017layer,zhang2017strong}
Furthermore, the non commensuration of graphene with the substrate creates Moire
patterns, resulting in an effective spatial modulation of the different
contributions.\cite{PhysRevB.90.075428,PhysRevB.96.085442,wang2016gaps}

%~~~~~~~~~~~~~~~~~~~~~~~~~~ FIGURE ~~~~~~~~~~~~~~~~~~~~~~~~~%
\begin{figure}[t!]
\centering
\includegraphics[width=0.5\textwidth]{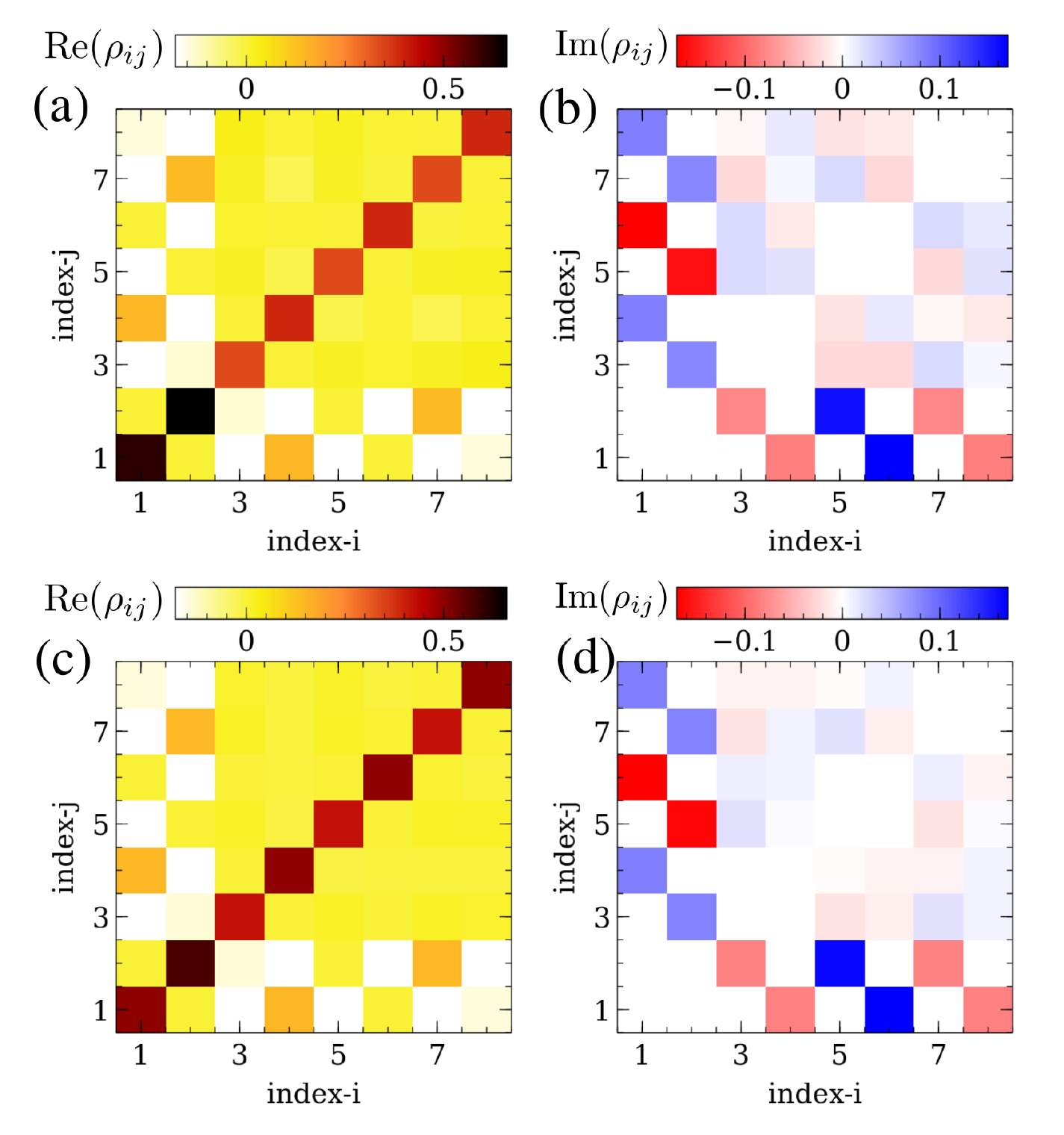}
\caption{
Real and imaginary parts of the density matrix for a trivial $\mathcal{C}=0$ (a,b)
and a topological $\mathcal{C}=2$ (c,d) two dimensional system.
In this case, the neural network will implement an image recognition algorithm,
where the input are the two images representing the real and imaginary parts.
}
\label{fig5}
\end{figure}
%~~~~~~~~~~~~~~~~~~~~~~~~~~~~~~~~~~~~~~~~~~~~~~~~~~~~~~~~~~~%

It is worth mentioning two important differences with respect to the model
presented in Sec.~\ref{sec:1d}. On one hand, now the Hamiltonian involves the
three Pauli matrices, so in general it will be complex. This implies that the
calculated density matrices will also be complex, so that the neural network
will receive as input both the real and imaginary components.
On the other hand, since we are dealing now with a two dimensional system, a
finite island will have $L^2$ sites, with $L$ the typical size of the island.
In particular, the calculation of a the density matrix with the wavefunctions of
an island with side $L\approx 300$ would require the diagonalization of matrix
of dimension $L^2\approx 90000$, whose computational complexity is $L^6$. It is
in this situation where the KPM will be specially useful, as it allows us to
calculate the density matrix with a computational complexity of the number of
sites, $L^2$.

%~~~~~~~~~~~~~~~~~~~~~~~~~~ FIGURE ~~~~~~~~~~~~~~~~~~~~~~~~~%
\begin{figure}[t!]
\centering
\includegraphics[width=0.5\textwidth]{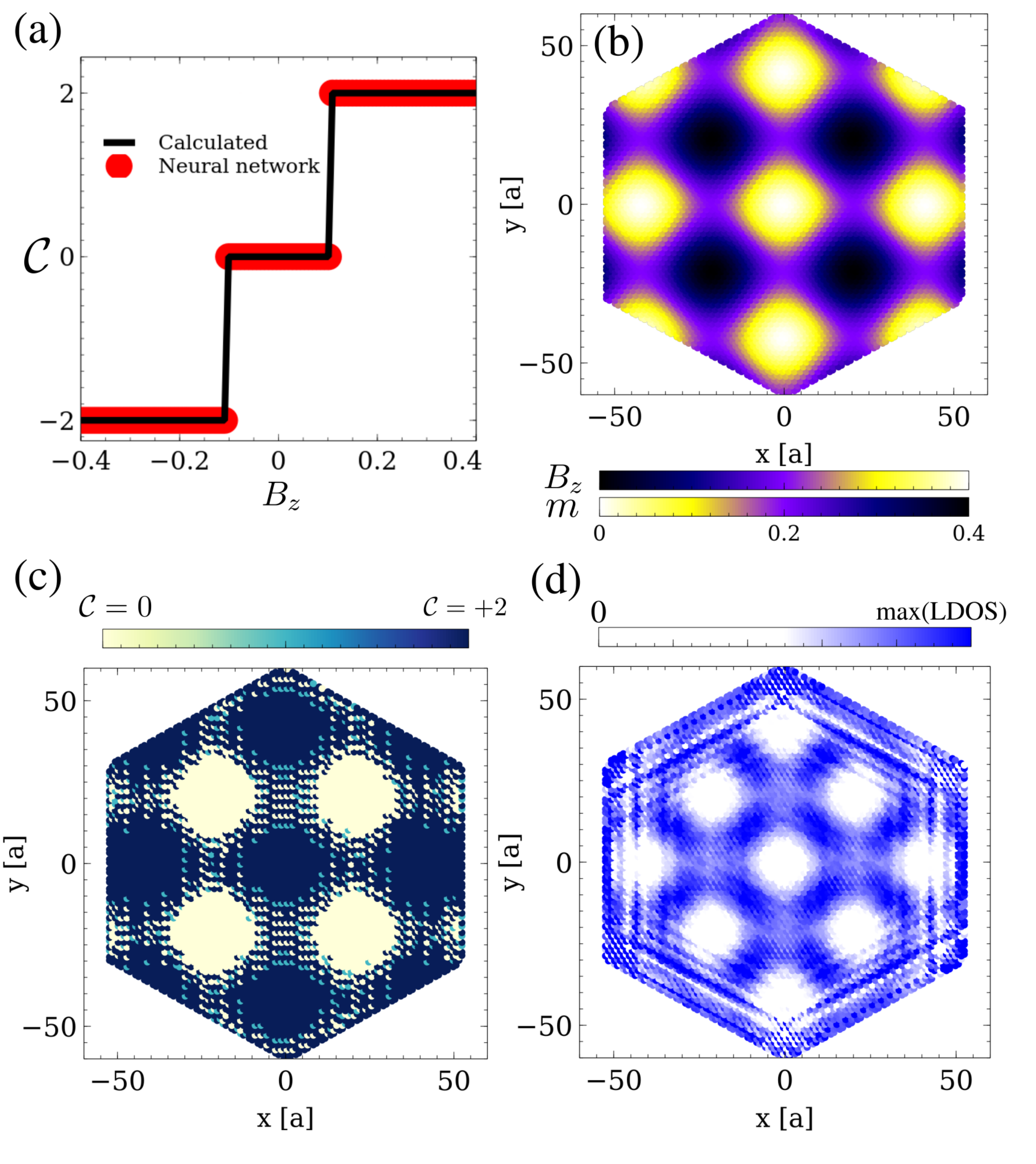}
\caption{
(a) Comparison between the exact Chern number (black) and the prediction of the
trained neural network (red) using as input the local density matrix. Once the
accuracy of the network has been checked, we created a big graphene island with
modulated mass and exchange term as shown in (b). The neural network is used
to evaluate the topological invariant in each atom, yielding the result shown
in (c). The boundary between different topological phases is expected to give
rise to in-gap states, which is confirmed by calculating the in-gap spectral
function as shown in (d).
}
\label{DOS_fig}
\label{fig6}
\end{figure}
%~~~~~~~~~~~~~~~~~~~~~~~~~~~~~~~~~~~~~~~~~~~~~~~~~~~~~~~~~~~%

We now move to apply our methodology to the system defined by
eq.~\eqref{hamil:2d}.
First, to train the neural network, we generate different spatially uniform
Hamiltonians by choosing randomly each of the coupling parameters. The Zeeman
and Rashba were randomly generated in the interval $t_R\in[-0.4t,0.4t]$ and the mass
between $m\in[0,0.4t]$. Again, random Anderson-like disorder comparable to the other
interactions are introduced all across the system $W_i\in [0.0,0.4t]$.
The training dataset consisted in 564 samples. 
% and the testing dataset had 586 samples.
For every set of parameters, we built the Hamiltonian as in eq.~\eqref{hamil:2d}
and calculated the local density matrix for the central atom and its three first
neighbors which are used as input of the network, in this case a 128-dimensional
array.
Again we chose having two hidden layers with 101 and 21 neurons.
It is worth considering again the challenging task of distinguishing
between different inputs as those shown in Fig.~\ref{fig5}, which
highlights that the classification of topological and trivial phases based only
in local properties is far from being a trivial task.

The output for each input was obtained by calculating the Chern number of the
ground state of the system integrating the Berry curvature in the Brillouin zone
of a translational invariant ($W_i=0$) Hamiltonian with the same parameters.
Once the network was trained, we tested its  accuracy on a validation
dataset with 586 samples randomly generated,  showing an accuracy of $\sim92\%$.
The comparison of the result predicted by the network and the one calculated
exactly in  a system with translational invariance is shown in
Fig.~\ref{fig6}~(a) for the different topological phases.
%Joaquin: Obvious questions here, and also in the 1D case: what is the origin of the errors. Is this related to Anderson disorder, or is a typical error in NN methods
%Noel: The errors are just borderline cases. An ANN will always smooth any abrupt change in the ouput, and in this case we are switching among 3 different cases which are discrete (no in-between). With more and more training you can reach better defined transitions, but if you zoom-in enough you will always find "errors" in the transition regions.

After the training, we generated a graphene nano-island with 7400 atoms. In this
island, we choose a spatially modulated exchange field of the form
$B_z(x,y) = 0.1t[\cos(0.15 x)+\cos(0.15 y)+2]$, a modulated mass term of the form
$m(x,y) = 0.1t[\sin(0.15 x)+\sin(0.15 y)+2]$ (shown in Fig.~\ref{fig6}~(b)), and a
constant Rashba coupling $\lambda_R = 0.2$.
The previous modulations are expected to create neighboring trivial and
topological areas depending on which is the dominant contribution, mass or
exchange and Rashba couplings.
With such a Hamiltonian, we calculated the local density matrix using the Kernel
polynomial method, that was used as input of the neural network.
The result of the evaluation of the neural network across the sample is shown in
Fig.~\ref{fig6}~(c). % Note that the transition regions are somewhat troublesome for the ANN. This is due to the fact that these algorithms provide a smooth output, so it would require much more training to sharpen the transitions.
It can be seen that different regions with different Chern number appear according
to the spatial modulation of the Hamiltonian parameters.
The significance of the different regions becomes clear once the in-gap density
of states is calculated in Fig.~\ref{fig6}~(d). This shows both in-gap modes
precisely at the boundary between different regions, as expected form the
bulk-boundary correspondence,  as well as edge states all around the sample.

This result  highlights that the artificial neural network faithfully distinguishes
between the different phases based solely in local information, providing an
useful method to calculate the topological invariant in systems without
translational symmetry.

%%%%%%%%%%%%%%%%%%%%%%%%%%%%%%%%%%%%%%%%%%%%%%%%%%%%%%%%%%%%%%%%%%%%%%%%%%%%%%%%
\section{Conclusions}
\label{sec:Conc}
We have shown that an artificial neural network is capable of predicting the
topological nature of different model Hamiltonians using as an input a local
sector of the density matrix, {\em i.e.}, evaluating solely \emph{local properties}.
Our procedure consisted on training an artificial neural network using as
input the subspace of the density matrix corresponding to a local area of the
sample, and as output the topological invariant that an analogous (pristine and
translational invariant) Hamiltonian with the same effective parameters would
have.

We applied this procedure to two well known models,  a  1D
topological superconductor and 2D topological insulator. In both
cases we considered finite systems with a space dependent Hamiltonian that
create regions with both topological and trivial character.
By evaluating the network with local quantities for each Hamiltonian we showed
that the different topological domains are accurately identified by the network,
even when the inhomogeneus systems have Anderson-like disorder, proving that
this methodology can be applied for disordered systems.

It is worth remarking that the training procedure is carried out for a specific
model, and tested in that same model for different parameters, including local
modulations in space.
An open question is whether this methodology can be extended to cases with the
same topological classes but different geometries.
%Joaquin: not sure of what this means. Examples?
%Noel: This means for instance that we are not sure if it should work or not in another QASH that is realized in any other material that is not a honeycomb lattice
Finally, it is interesting to note that an analogous methodology could be
applied  
to interacting systems, so that similar procedures could be exploited
to identify
quantum spin liquid states in two dimensional spin systems.

\section*{Acknowledgments}
This work has been financially supported in part by
FEDER funds. We acknowledge financial support by
Marie-Curie-ITN 607904-SPINOGRAPH , FCT, under
the project PTDC/FIS-NAN/4662/2014,
P2020-PTDC/FIS-NAN/3668/2014, and MINECO-Spain
(MAT2016-78625-C2). J.L. Lado
acknowledges financial support from ETH Fellowship program.
D. Carvalho acknowledges
the hospitality of INL through its INL Summer Student program.

%~~~~~~~~ Bibliography ~~~~~~~~
%\bibliographystyle{ieeetr}
\bibliographystyle{apsrev4-1}
\bibliography{biblio}

%merlin.mbs apsrev4-1.bst 2010-07-25 4.21a (PWD, AO, DPC) hacked
%Control: key (0)
%Control: author (72) initials jnrlst
%Control: editor formatted (1) identically to author
%Control: production of article title (-1) disabled
%Control: page (0) single
%Control: year (1) truncated
%Control: production of eprint (0) enabled
\begin{thebibliography}{66}%
\makeatletter
\providecommand \@ifxundefined [1]{%
 \@ifx{#1\undefined}
}%
\providecommand \@ifnum [1]{%
 \ifnum #1\expandafter \@firstoftwo
 \else \expandafter \@secondoftwo
 \fi
}%
\providecommand \@ifx [1]{%
 \ifx #1\expandafter \@firstoftwo
 \else \expandafter \@secondoftwo
 \fi
}%
\providecommand \natexlab [1]{#1}%
\providecommand \enquote  [1]{``#1''}%
\providecommand \bibnamefont  [1]{#1}%
\providecommand \bibfnamefont [1]{#1}%
\providecommand \citenamefont [1]{#1}%
\providecommand \href@noop [0]{\@secondoftwo}%
\providecommand \href [0]{\begingroup \@sanitize@url \@href}%
\providecommand \@href[1]{\@@startlink{#1}\@@href}%
\providecommand \@@href[1]{\endgroup#1\@@endlink}%
\providecommand \@sanitize@url [0]{\catcode `\\12\catcode `\$12\catcode
  `\&12\catcode `\#12\catcode `\^12\catcode `\_12\catcode `\%12\relax}%
\providecommand \@@startlink[1]{}%
\providecommand \@@endlink[0]{}%
\providecommand \url  [0]{\begingroup\@sanitize@url \@url }%
\providecommand \@url [1]{\endgroup\@href {#1}{\urlprefix }}%
\providecommand \urlprefix  [0]{URL }%
\providecommand \Eprint [0]{\href }%
\providecommand \doibase [0]{http://dx.doi.org/}%
\providecommand \selectlanguage [0]{\@gobble}%
\providecommand \bibinfo  [0]{\@secondoftwo}%
\providecommand \bibfield  [0]{\@secondoftwo}%
\providecommand \translation [1]{[#1]}%
\providecommand \BibitemOpen [0]{}%
\providecommand \bibitemStop [0]{}%
\providecommand \bibitemNoStop [0]{.\EOS\space}%
\providecommand \EOS [0]{\spacefactor3000\relax}%
\providecommand \BibitemShut  [1]{\csname bibitem#1\endcsname}%
\let\auto@bib@innerbib\@empty
%</preamble>
\bibitem [{\citenamefont {Hasan}\ and\ \citenamefont
  {Kane}(2010)}]{RevModPhys.82.3045}%
  \BibitemOpen
  \bibfield  {author} {\bibinfo {author} {\bibfnamefont {M.~Z.}\ \bibnamefont
  {Hasan}}\ and\ \bibinfo {author} {\bibfnamefont {C.~L.}\ \bibnamefont
  {Kane}},\ }\href {\doibase 10.1103/RevModPhys.82.3045} {\bibfield  {journal}
  {\bibinfo  {journal} {Rev. Mod. Phys.}\ }\textbf {\bibinfo {volume} {82}},\
  \bibinfo {pages} {3045} (\bibinfo {year} {2010})}\BibitemShut {NoStop}%
\bibitem [{\citenamefont {Qi}\ and\ \citenamefont
  {Zhang}(2011)}]{RevModPhys.83.1057}%
  \BibitemOpen
  \bibfield  {author} {\bibinfo {author} {\bibfnamefont {X.-L.}\ \bibnamefont
  {Qi}}\ and\ \bibinfo {author} {\bibfnamefont {S.-C.}\ \bibnamefont {Zhang}},\
  }\href {\doibase 10.1103/RevModPhys.83.1057} {\bibfield  {journal} {\bibinfo
  {journal} {Rev. Mod. Phys.}\ }\textbf {\bibinfo {volume} {83}},\ \bibinfo
  {pages} {1057} (\bibinfo {year} {2011})}\BibitemShut {NoStop}%
\bibitem [{\citenamefont {Fu}(2011)}]{PhysRevLett.106.106802}%
  \BibitemOpen
  \bibfield  {author} {\bibinfo {author} {\bibfnamefont {L.}~\bibnamefont
  {Fu}},\ }\href {\doibase 10.1103/PhysRevLett.106.106802} {\bibfield
  {journal} {\bibinfo  {journal} {Phys. Rev. Lett.}\ }\textbf {\bibinfo
  {volume} {106}},\ \bibinfo {pages} {106802} (\bibinfo {year}
  {2011})}\BibitemShut {NoStop}%
\bibitem [{\citenamefont {Dzero}\ \emph {et~al.}(2010)\citenamefont {Dzero},
  \citenamefont {Sun}, \citenamefont {Galitski},\ and\ \citenamefont
  {Coleman}}]{PhysRevLett.104.106408}%
  \BibitemOpen
  \bibfield  {author} {\bibinfo {author} {\bibfnamefont {M.}~\bibnamefont
  {Dzero}}, \bibinfo {author} {\bibfnamefont {K.}~\bibnamefont {Sun}}, \bibinfo
  {author} {\bibfnamefont {V.}~\bibnamefont {Galitski}}, \ and\ \bibinfo
  {author} {\bibfnamefont {P.}~\bibnamefont {Coleman}},\ }\href {\doibase
  10.1103/PhysRevLett.104.106408} {\bibfield  {journal} {\bibinfo  {journal}
  {Phys. Rev. Lett.}\ }\textbf {\bibinfo {volume} {104}},\ \bibinfo {pages}
  {106408} (\bibinfo {year} {2010})}\BibitemShut {NoStop}%
\bibitem [{\citenamefont {Raghu}\ \emph {et~al.}(2008)\citenamefont {Raghu},
  \citenamefont {Qi}, \citenamefont {Honerkamp},\ and\ \citenamefont
  {Zhang}}]{PhysRevLett.100.156401}%
  \BibitemOpen
  \bibfield  {author} {\bibinfo {author} {\bibfnamefont {S.}~\bibnamefont
  {Raghu}}, \bibinfo {author} {\bibfnamefont {X.-L.}\ \bibnamefont {Qi}},
  \bibinfo {author} {\bibfnamefont {C.}~\bibnamefont {Honerkamp}}, \ and\
  \bibinfo {author} {\bibfnamefont {S.-C.}\ \bibnamefont {Zhang}},\ }\href
  {\doibase 10.1103/PhysRevLett.100.156401} {\bibfield  {journal} {\bibinfo
  {journal} {Phys. Rev. Lett.}\ }\textbf {\bibinfo {volume} {100}},\ \bibinfo
  {pages} {156401} (\bibinfo {year} {2008})}\BibitemShut {NoStop}%
\bibitem [{\citenamefont {Soluyanov}\ and\ \citenamefont
  {Vanderbilt}(2011)}]{PhysRevB.83.235401}%
  \BibitemOpen
  \bibfield  {author} {\bibinfo {author} {\bibfnamefont {A.~A.}\ \bibnamefont
  {Soluyanov}}\ and\ \bibinfo {author} {\bibfnamefont {D.}~\bibnamefont
  {Vanderbilt}},\ }\href {\doibase 10.1103/PhysRevB.83.235401} {\bibfield
  {journal} {\bibinfo  {journal} {Phys. Rev. B}\ }\textbf {\bibinfo {volume}
  {83}},\ \bibinfo {pages} {235401} (\bibinfo {year} {2011})}\BibitemShut
  {NoStop}%
\bibitem [{\citenamefont {Gresch}\ \emph {et~al.}(2017)\citenamefont {Gresch},
  \citenamefont {Aut\`es}, \citenamefont {Yazyev}, \citenamefont {Troyer},
  \citenamefont {Vanderbilt}, \citenamefont {Bernevig},\ and\ \citenamefont
  {Soluyanov}}]{PhysRevB.95.075146}%
  \BibitemOpen
  \bibfield  {author} {\bibinfo {author} {\bibfnamefont {D.}~\bibnamefont
  {Gresch}}, \bibinfo {author} {\bibfnamefont {G.}~\bibnamefont {Aut\`es}},
  \bibinfo {author} {\bibfnamefont {O.~V.}\ \bibnamefont {Yazyev}}, \bibinfo
  {author} {\bibfnamefont {M.}~\bibnamefont {Troyer}}, \bibinfo {author}
  {\bibfnamefont {D.}~\bibnamefont {Vanderbilt}}, \bibinfo {author}
  {\bibfnamefont {B.~A.}\ \bibnamefont {Bernevig}}, \ and\ \bibinfo {author}
  {\bibfnamefont {A.~A.}\ \bibnamefont {Soluyanov}},\ }\href {\doibase
  10.1103/PhysRevB.95.075146} {\bibfield  {journal} {\bibinfo  {journal} {Phys.
  Rev. B}\ }\textbf {\bibinfo {volume} {95}},\ \bibinfo {pages} {075146}
  (\bibinfo {year} {2017})}\BibitemShut {NoStop}%
\bibitem [{\citenamefont {Wu}\ \emph {et~al.}(2017)\citenamefont {Wu},
  \citenamefont {Zhang}, \citenamefont {Song}, \citenamefont {Troyer},\ and\
  \citenamefont {Soluyanov}}]{wu2017wanniertools}%
  \BibitemOpen
  \bibfield  {author} {\bibinfo {author} {\bibfnamefont {Q.}~\bibnamefont
  {Wu}}, \bibinfo {author} {\bibfnamefont {S.}~\bibnamefont {Zhang}}, \bibinfo
  {author} {\bibfnamefont {H.-F.}\ \bibnamefont {Song}}, \bibinfo {author}
  {\bibfnamefont {M.}~\bibnamefont {Troyer}}, \ and\ \bibinfo {author}
  {\bibfnamefont {A.~A.}\ \bibnamefont {Soluyanov}},\ }\href@noop {} {\bibfield
   {journal} {\bibinfo  {journal} {arXiv preprint arXiv:1703.07789}\ }
  (\bibinfo {year} {2017})}\BibitemShut {NoStop}%
\bibitem [{\citenamefont {San-Jose}\ \emph {et~al.}(2014)\citenamefont
  {San-Jose}, \citenamefont {Guti\'errez-Rubio}, \citenamefont {Sturla},\ and\
  \citenamefont {Guinea}}]{PhysRevB.90.075428}%
  \BibitemOpen
  \bibfield  {author} {\bibinfo {author} {\bibfnamefont {P.}~\bibnamefont
  {San-Jose}}, \bibinfo {author} {\bibfnamefont {A.}~\bibnamefont
  {Guti\'errez-Rubio}}, \bibinfo {author} {\bibfnamefont {M.}~\bibnamefont
  {Sturla}}, \ and\ \bibinfo {author} {\bibfnamefont {F.}~\bibnamefont
  {Guinea}},\ }\href {\doibase 10.1103/PhysRevB.90.075428} {\bibfield
  {journal} {\bibinfo  {journal} {Phys. Rev. B}\ }\textbf {\bibinfo {volume}
  {90}},\ \bibinfo {pages} {075428} (\bibinfo {year} {2014})}\BibitemShut
  {NoStop}%
\bibitem [{\citenamefont {Jung}\ \emph {et~al.}(2017)\citenamefont {Jung},
  \citenamefont {Laksono}, \citenamefont {DaSilva}, \citenamefont {MacDonald},
  \citenamefont {Mucha-Kruczy\ifmmode~\acute{n}\else \'{n}\fi{}ski},\ and\
  \citenamefont {Adam}}]{PhysRevB.96.085442}%
  \BibitemOpen
  \bibfield  {author} {\bibinfo {author} {\bibfnamefont {J.}~\bibnamefont
  {Jung}}, \bibinfo {author} {\bibfnamefont {E.}~\bibnamefont {Laksono}},
  \bibinfo {author} {\bibfnamefont {A.~M.}\ \bibnamefont {DaSilva}}, \bibinfo
  {author} {\bibfnamefont {A.~H.}\ \bibnamefont {MacDonald}}, \bibinfo {author}
  {\bibfnamefont {M.}~\bibnamefont {Mucha-Kruczy\ifmmode~\acute{n}\else
  \'{n}\fi{}ski}}, \ and\ \bibinfo {author} {\bibfnamefont {S.}~\bibnamefont
  {Adam}},\ }\href {\doibase 10.1103/PhysRevB.96.085442} {\bibfield  {journal}
  {\bibinfo  {journal} {Phys. Rev. B}\ }\textbf {\bibinfo {volume} {96}},\
  \bibinfo {pages} {085442} (\bibinfo {year} {2017})}\BibitemShut {NoStop}%
\bibitem [{\citenamefont {Wang}\ \emph {et~al.}(2016)\citenamefont {Wang},
  \citenamefont {Lu}, \citenamefont {Ding}, \citenamefont {Yao}, \citenamefont
  {Yan}, \citenamefont {Wan}, \citenamefont {Deng}, \citenamefont {Wang},
  \citenamefont {Chen}, \citenamefont {Ma} \emph {et~al.}}]{wang2016gaps}%
  \BibitemOpen
  \bibfield  {author} {\bibinfo {author} {\bibfnamefont {E.}~\bibnamefont
  {Wang}}, \bibinfo {author} {\bibfnamefont {X.}~\bibnamefont {Lu}}, \bibinfo
  {author} {\bibfnamefont {S.}~\bibnamefont {Ding}}, \bibinfo {author}
  {\bibfnamefont {W.}~\bibnamefont {Yao}}, \bibinfo {author} {\bibfnamefont
  {M.}~\bibnamefont {Yan}}, \bibinfo {author} {\bibfnamefont {G.}~\bibnamefont
  {Wan}}, \bibinfo {author} {\bibfnamefont {K.}~\bibnamefont {Deng}}, \bibinfo
  {author} {\bibfnamefont {S.}~\bibnamefont {Wang}}, \bibinfo {author}
  {\bibfnamefont {G.}~\bibnamefont {Chen}}, \bibinfo {author} {\bibfnamefont
  {L.}~\bibnamefont {Ma}},  \emph {et~al.},\ }\href@noop {} {\bibfield
  {journal} {\bibinfo  {journal} {Nature Physics}\ }\textbf {\bibinfo {volume}
  {12}},\ \bibinfo {pages} {1111} (\bibinfo {year} {2016})}\BibitemShut
  {NoStop}%
\bibitem [{\citenamefont {Sanchez-Yamagishi}\ \emph {et~al.}(2017)\citenamefont
  {Sanchez-Yamagishi}, \citenamefont {Luo}, \citenamefont {Young},
  \citenamefont {Hunt}, \citenamefont {Watanabe}, \citenamefont {Taniguchi},
  \citenamefont {Ashoori},\ and\ \citenamefont
  {Jarillo-Herrero}}]{sanchez2017helical}%
  \BibitemOpen
  \bibfield  {author} {\bibinfo {author} {\bibfnamefont {J.~D.}\ \bibnamefont
  {Sanchez-Yamagishi}}, \bibinfo {author} {\bibfnamefont {J.~Y.}\ \bibnamefont
  {Luo}}, \bibinfo {author} {\bibfnamefont {A.~F.}\ \bibnamefont {Young}},
  \bibinfo {author} {\bibfnamefont {B.~M.}\ \bibnamefont {Hunt}}, \bibinfo
  {author} {\bibfnamefont {K.}~\bibnamefont {Watanabe}}, \bibinfo {author}
  {\bibfnamefont {T.}~\bibnamefont {Taniguchi}}, \bibinfo {author}
  {\bibfnamefont {R.~C.}\ \bibnamefont {Ashoori}}, \ and\ \bibinfo {author}
  {\bibfnamefont {P.}~\bibnamefont {Jarillo-Herrero}},\ }\href@noop {}
  {\bibfield  {journal} {\bibinfo  {journal} {Nature nanotechnology}\ }\textbf
  {\bibinfo {volume} {12}},\ \bibinfo {pages} {118} (\bibinfo {year}
  {2017})}\BibitemShut {NoStop}%
\bibitem [{\citenamefont {Young}\ \emph {et~al.}(2014)\citenamefont {Young},
  \citenamefont {Sanchez-Yamagishi}, \citenamefont {Hunt}, \citenamefont
  {Choi}, \citenamefont {Watanabe}, \citenamefont {Taniguchi}, \citenamefont
  {Ashoori},\ and\ \citenamefont {Jarillo-Herrero}}]{Young2014}%
  \BibitemOpen
  \bibfield  {author} {\bibinfo {author} {\bibfnamefont {A.~F.}\ \bibnamefont
  {Young}}, \bibinfo {author} {\bibfnamefont {J.~D.}\ \bibnamefont
  {Sanchez-Yamagishi}}, \bibinfo {author} {\bibfnamefont {B.}~\bibnamefont
  {Hunt}}, \bibinfo {author} {\bibfnamefont {S.~H.}\ \bibnamefont {Choi}},
  \bibinfo {author} {\bibfnamefont {K.}~\bibnamefont {Watanabe}}, \bibinfo
  {author} {\bibfnamefont {T.}~\bibnamefont {Taniguchi}}, \bibinfo {author}
  {\bibfnamefont {R.~C.}\ \bibnamefont {Ashoori}}, \ and\ \bibinfo {author}
  {\bibfnamefont {P.}~\bibnamefont {Jarillo-Herrero}},\ }\href
  {http://dx.doi.org/10.1038/nature12800} {\bibfield  {journal} {\bibinfo
  {journal} {Nature}\ }\textbf {\bibinfo {volume} {505}},\ \bibinfo {pages}
  {528} (\bibinfo {year} {2014})},\ \bibinfo {note} {letter}\BibitemShut
  {NoStop}%
\bibitem [{\citenamefont {Alicea}\ \emph {et~al.}(2011)\citenamefont {Alicea},
  \citenamefont {Oreg}, \citenamefont {Refael}, \citenamefont {von Oppen},\
  and\ \citenamefont {Fisher}}]{Alicea2011}%
  \BibitemOpen
  \bibfield  {author} {\bibinfo {author} {\bibfnamefont {J.}~\bibnamefont
  {Alicea}}, \bibinfo {author} {\bibfnamefont {Y.}~\bibnamefont {Oreg}},
  \bibinfo {author} {\bibfnamefont {G.}~\bibnamefont {Refael}}, \bibinfo
  {author} {\bibfnamefont {F.}~\bibnamefont {von Oppen}}, \ and\ \bibinfo
  {author} {\bibfnamefont {M.~P.~A.}\ \bibnamefont {Fisher}},\ }\href
  {http://dx.doi.org/10.1038/nphys1915} {\ \textbf {\bibinfo {volume} {7}},\
  \bibinfo {pages} {412 EP } (\bibinfo {year} {2011})},\ \bibinfo {note}
  {article}\BibitemShut {NoStop}%
\bibitem [{\citenamefont {Zhang}\ \emph {et~al.}(2016)\citenamefont {Zhang},
  \citenamefont {G{\"u}l}, \citenamefont {Conesa-Boj}, \citenamefont {Zuo},
  \citenamefont {Mourik}, \citenamefont {de~Vries}, \citenamefont {van Veen},
  \citenamefont {van Woerkom}, \citenamefont {Nowak}, \citenamefont {Wimmer}
  \emph {et~al.}}]{zhang2016ballistic}%
  \BibitemOpen
  \bibfield  {author} {\bibinfo {author} {\bibfnamefont {H.}~\bibnamefont
  {Zhang}}, \bibinfo {author} {\bibfnamefont {{\"O}.}~\bibnamefont {G{\"u}l}},
  \bibinfo {author} {\bibfnamefont {S.}~\bibnamefont {Conesa-Boj}}, \bibinfo
  {author} {\bibfnamefont {K.}~\bibnamefont {Zuo}}, \bibinfo {author}
  {\bibfnamefont {V.}~\bibnamefont {Mourik}}, \bibinfo {author} {\bibfnamefont
  {F.~K.}\ \bibnamefont {de~Vries}}, \bibinfo {author} {\bibfnamefont
  {J.}~\bibnamefont {van Veen}}, \bibinfo {author} {\bibfnamefont {D.~J.}\
  \bibnamefont {van Woerkom}}, \bibinfo {author} {\bibfnamefont {M.~P.}\
  \bibnamefont {Nowak}}, \bibinfo {author} {\bibfnamefont {M.}~\bibnamefont
  {Wimmer}},  \emph {et~al.},\ }\href@noop {} {\bibfield  {journal} {\bibinfo
  {journal} {arXiv preprint arXiv:1603.04069}\ } (\bibinfo {year}
  {2016})}\BibitemShut {NoStop}%
\bibitem [{\citenamefont {Bianco}\ and\ \citenamefont
  {Resta}(2011)}]{PhysRevB.84.241106}%
  \BibitemOpen
  \bibfield  {author} {\bibinfo {author} {\bibfnamefont {R.}~\bibnamefont
  {Bianco}}\ and\ \bibinfo {author} {\bibfnamefont {R.}~\bibnamefont {Resta}},\
  }\href {\doibase 10.1103/PhysRevB.84.241106} {\bibfield  {journal} {\bibinfo
  {journal} {Phys. Rev. B}\ }\textbf {\bibinfo {volume} {84}},\ \bibinfo
  {pages} {241106} (\bibinfo {year} {2011})}\BibitemShut {NoStop}%
\bibitem [{\citenamefont {Marrazzo}\ and\ \citenamefont
  {Resta}(2017)}]{PhysRevB.95.121114}%
  \BibitemOpen
  \bibfield  {author} {\bibinfo {author} {\bibfnamefont {A.}~\bibnamefont
  {Marrazzo}}\ and\ \bibinfo {author} {\bibfnamefont {R.}~\bibnamefont
  {Resta}},\ }\href {\doibase 10.1103/PhysRevB.95.121114} {\bibfield  {journal}
  {\bibinfo  {journal} {Phys. Rev. B}\ }\textbf {\bibinfo {volume} {95}},\
  \bibinfo {pages} {121114} (\bibinfo {year} {2017})}\BibitemShut {NoStop}%
\bibitem [{\citenamefont {Loring}(2015)}]{loring2015k}%
  \BibitemOpen
  \bibfield  {author} {\bibinfo {author} {\bibfnamefont {T.~A.}\ \bibnamefont
  {Loring}},\ }\href@noop {} {\bibfield  {journal} {\bibinfo  {journal} {Annals
  of Physics}\ }\textbf {\bibinfo {volume} {356}},\ \bibinfo {pages} {383}
  (\bibinfo {year} {2015})}\BibitemShut {NoStop}%
\bibitem [{\citenamefont {Mitchell}\ \emph {et~al.}(2018)\citenamefont
  {Mitchell}, \citenamefont {Nash}, \citenamefont {Hexner}, \citenamefont
  {Turner},\ and\ \citenamefont {Irvine}}]{mitchell2018amorphous}%
  \BibitemOpen
  \bibfield  {author} {\bibinfo {author} {\bibfnamefont {N.~P.}\ \bibnamefont
  {Mitchell}}, \bibinfo {author} {\bibfnamefont {L.~M.}\ \bibnamefont {Nash}},
  \bibinfo {author} {\bibfnamefont {D.}~\bibnamefont {Hexner}}, \bibinfo
  {author} {\bibfnamefont {A.~M.}\ \bibnamefont {Turner}}, \ and\ \bibinfo
  {author} {\bibfnamefont {W.~T.}\ \bibnamefont {Irvine}},\ }\href@noop {}
  {\bibfield  {journal} {\bibinfo  {journal} {Nature Physics}\ ,\ \bibinfo
  {pages} {1}} (\bibinfo {year} {2018})}\BibitemShut {NoStop}%
\bibitem [{\citenamefont {Fulga}\ \emph {et~al.}(2016)\citenamefont {Fulga},
  \citenamefont {Pikulin},\ and\ \citenamefont {Loring}}]{fulga2016aperiodic}%
  \BibitemOpen
  \bibfield  {author} {\bibinfo {author} {\bibfnamefont {I.~C.}\ \bibnamefont
  {Fulga}}, \bibinfo {author} {\bibfnamefont {D.~I.}\ \bibnamefont {Pikulin}},
  \ and\ \bibinfo {author} {\bibfnamefont {T.~A.}\ \bibnamefont {Loring}},\
  }\href@noop {} {\bibfield  {journal} {\bibinfo  {journal} {Physical review
  letters}\ }\textbf {\bibinfo {volume} {116}},\ \bibinfo {pages} {257002}
  (\bibinfo {year} {2016})}\BibitemShut {NoStop}%
\bibitem [{\citenamefont {van Nieuwenburg}\ \emph {et~al.}(2017)\citenamefont
  {van Nieuwenburg}, \citenamefont {Liu},\ and\ \citenamefont
  {Huber}}]{VanNieuwenburg2017}%
  \BibitemOpen
  \bibfield  {author} {\bibinfo {author} {\bibfnamefont {E.}~\bibnamefont {van
  Nieuwenburg}}, \bibinfo {author} {\bibfnamefont {Y.-H.}\ \bibnamefont {Liu}},
  \ and\ \bibinfo {author} {\bibfnamefont {S.}~\bibnamefont {Huber}},\ }\href
  {\doibase 10.1038/nphys4037} {\bibfield  {journal} {\bibinfo  {journal}
  {Nature Physics}\ }\textbf {\bibinfo {volume} {13}},\ \bibinfo {pages} {435}
  (\bibinfo {year} {2017})},\ \Eprint {http://arxiv.org/abs/arXiv:1610.02048}
  {arXiv:1610.02048} \BibitemShut {NoStop}%
\bibitem [{\citenamefont {Ch'ng}\ \emph {et~al.}(2017)\citenamefont {Ch'ng},
  \citenamefont {Carrasquilla}, \citenamefont {Melko},\ and\ \citenamefont
  {Khatami}}]{PhysRevX.7.031038}%
  \BibitemOpen
  \bibfield  {author} {\bibinfo {author} {\bibfnamefont {K.}~\bibnamefont
  {Ch'ng}}, \bibinfo {author} {\bibfnamefont {J.}~\bibnamefont {Carrasquilla}},
  \bibinfo {author} {\bibfnamefont {R.~G.}\ \bibnamefont {Melko}}, \ and\
  \bibinfo {author} {\bibfnamefont {E.}~\bibnamefont {Khatami}},\ }\href
  {\doibase 10.1103/PhysRevX.7.031038} {\bibfield  {journal} {\bibinfo
  {journal} {Phys. Rev. X}\ }\textbf {\bibinfo {volume} {7}},\ \bibinfo {pages}
  {031038} (\bibinfo {year} {2017})}\BibitemShut {NoStop}%
\bibitem [{\citenamefont {Ohtsuki}\ and\ \citenamefont
  {Ohtsuki}(2016)}]{ohtsuki2016deep}%
  \BibitemOpen
  \bibfield  {author} {\bibinfo {author} {\bibfnamefont {T.}~\bibnamefont
  {Ohtsuki}}\ and\ \bibinfo {author} {\bibfnamefont {T.}~\bibnamefont
  {Ohtsuki}},\ }\href@noop {} {\bibfield  {journal} {\bibinfo  {journal}
  {Journal of the Physical Society of Japan}\ }\textbf {\bibinfo {volume}
  {85}},\ \bibinfo {pages} {123706} (\bibinfo {year} {2016})}\BibitemShut
  {NoStop}%
\bibitem [{\citenamefont {Hu}\ \emph {et~al.}(2017)\citenamefont {Hu},
  \citenamefont {Singh},\ and\ \citenamefont {Scalettar}}]{PhysRevE.95.062122}%
  \BibitemOpen
  \bibfield  {author} {\bibinfo {author} {\bibfnamefont {W.}~\bibnamefont
  {Hu}}, \bibinfo {author} {\bibfnamefont {R.~R.~P.}\ \bibnamefont {Singh}}, \
  and\ \bibinfo {author} {\bibfnamefont {R.~T.}\ \bibnamefont {Scalettar}},\
  }\href {\doibase 10.1103/PhysRevE.95.062122} {\bibfield  {journal} {\bibinfo
  {journal} {Phys. Rev. E}\ }\textbf {\bibinfo {volume} {95}},\ \bibinfo
  {pages} {062122} (\bibinfo {year} {2017})}\BibitemShut {NoStop}%
\bibitem [{\citenamefont {Broecker}\ \emph {et~al.}(2017)\citenamefont
  {Broecker}, \citenamefont {Assaad},\ and\ \citenamefont
  {Trebst}}]{broecker2017quantum}%
  \BibitemOpen
  \bibfield  {author} {\bibinfo {author} {\bibfnamefont {P.}~\bibnamefont
  {Broecker}}, \bibinfo {author} {\bibfnamefont {F.~F.}\ \bibnamefont
  {Assaad}}, \ and\ \bibinfo {author} {\bibfnamefont {S.}~\bibnamefont
  {Trebst}},\ }\href@noop {} {\bibfield  {journal} {\bibinfo  {journal} {arXiv
  preprint arXiv:1707.00663}\ } (\bibinfo {year} {2017})}\BibitemShut {NoStop}%
\bibitem [{\citenamefont {Koch-Janusz}\ and\ \citenamefont
  {Ringel}(2017)}]{koch2017mutual}%
  \BibitemOpen
  \bibfield  {author} {\bibinfo {author} {\bibfnamefont {M.}~\bibnamefont
  {Koch-Janusz}}\ and\ \bibinfo {author} {\bibfnamefont {Z.}~\bibnamefont
  {Ringel}},\ }\href@noop {} {\bibfield  {journal} {\bibinfo  {journal} {arXiv
  preprint arXiv:1704.06279}\ } (\bibinfo {year} {2017})}\BibitemShut {NoStop}%
\bibitem [{\citenamefont {Carleo}\ and\ \citenamefont
  {Troyer}(2017)}]{carleo2017solving}%
  \BibitemOpen
  \bibfield  {author} {\bibinfo {author} {\bibfnamefont {G.}~\bibnamefont
  {Carleo}}\ and\ \bibinfo {author} {\bibfnamefont {M.}~\bibnamefont
  {Troyer}},\ }\href@noop {} {\bibfield  {journal} {\bibinfo  {journal}
  {Science}\ }\textbf {\bibinfo {volume} {355}},\ \bibinfo {pages} {602}
  (\bibinfo {year} {2017})}\BibitemShut {NoStop}%
\bibitem [{\citenamefont {Deng}\ \emph {et~al.}(2016)\citenamefont {Deng},
  \citenamefont {Li},\ and\ \citenamefont {Sarma}}]{deng2016exact}%
  \BibitemOpen
  \bibfield  {author} {\bibinfo {author} {\bibfnamefont {D.-L.}\ \bibnamefont
  {Deng}}, \bibinfo {author} {\bibfnamefont {X.}~\bibnamefont {Li}}, \ and\
  \bibinfo {author} {\bibfnamefont {S.~D.}\ \bibnamefont {Sarma}},\ }\href@noop
  {} {\bibfield  {journal} {\bibinfo  {journal} {arXiv preprint
  arXiv:1609.09060}\ } (\bibinfo {year} {2016})}\BibitemShut {NoStop}%
\bibitem [{\citenamefont {Deng}\ \emph {et~al.}(2017)\citenamefont {Deng},
  \citenamefont {Li},\ and\ \citenamefont {Das~Sarma}}]{PhysRevX.7.021021}%
  \BibitemOpen
  \bibfield  {author} {\bibinfo {author} {\bibfnamefont {D.-L.}\ \bibnamefont
  {Deng}}, \bibinfo {author} {\bibfnamefont {X.}~\bibnamefont {Li}}, \ and\
  \bibinfo {author} {\bibfnamefont {S.}~\bibnamefont {Das~Sarma}},\ }\href
  {\doibase 10.1103/PhysRevX.7.021021} {\bibfield  {journal} {\bibinfo
  {journal} {Phys. Rev. X}\ }\textbf {\bibinfo {volume} {7}},\ \bibinfo {pages}
  {021021} (\bibinfo {year} {2017})}\BibitemShut {NoStop}%
\bibitem [{\citenamefont {Zhang}\ and\ \citenamefont
  {Kim}(2017)}]{PhysRevLett.118.216401}%
  \BibitemOpen
  \bibfield  {author} {\bibinfo {author} {\bibfnamefont {Y.}~\bibnamefont
  {Zhang}}\ and\ \bibinfo {author} {\bibfnamefont {E.-A.}\ \bibnamefont
  {Kim}},\ }\href {\doibase 10.1103/PhysRevLett.118.216401} {\bibfield
  {journal} {\bibinfo  {journal} {Phys. Rev. Lett.}\ }\textbf {\bibinfo
  {volume} {118}},\ \bibinfo {pages} {216401} (\bibinfo {year}
  {2017})}\BibitemShut {NoStop}%
\bibitem [{\citenamefont {Nagai}\ \emph {et~al.}(2017)\citenamefont {Nagai},
  \citenamefont {Shen}, \citenamefont {Qi}, \citenamefont {Liu},\ and\
  \citenamefont {Fu}}]{nagai2017self}%
  \BibitemOpen
  \bibfield  {author} {\bibinfo {author} {\bibfnamefont {Y.}~\bibnamefont
  {Nagai}}, \bibinfo {author} {\bibfnamefont {H.}~\bibnamefont {Shen}},
  \bibinfo {author} {\bibfnamefont {Y.}~\bibnamefont {Qi}}, \bibinfo {author}
  {\bibfnamefont {J.}~\bibnamefont {Liu}}, \ and\ \bibinfo {author}
  {\bibfnamefont {L.}~\bibnamefont {Fu}},\ }\href {\doibase
  10.1103/PhysRevB.96.161102} {\bibfield  {journal} {\bibinfo  {journal} {Phys.
  Rev. B}\ }\textbf {\bibinfo {volume} {96}},\ \bibinfo {pages} {161102}
  (\bibinfo {year} {2017})}\BibitemShut {NoStop}%
\bibitem [{\citenamefont {Ch'ng}\ \emph {et~al.}(2018)\citenamefont {Ch'ng},
  \citenamefont {Vazquez},\ and\ \citenamefont {Khatami}}]{PhysRevE.97.013306}%
  \BibitemOpen
  \bibfield  {author} {\bibinfo {author} {\bibfnamefont {K.}~\bibnamefont
  {Ch'ng}}, \bibinfo {author} {\bibfnamefont {N.}~\bibnamefont {Vazquez}}, \
  and\ \bibinfo {author} {\bibfnamefont {E.}~\bibnamefont {Khatami}},\ }\href
  {\doibase 10.1103/PhysRevE.97.013306} {\bibfield  {journal} {\bibinfo
  {journal} {Phys. Rev. E}\ }\textbf {\bibinfo {volume} {97}},\ \bibinfo
  {pages} {013306} (\bibinfo {year} {2018})}\BibitemShut {NoStop}%
\bibitem [{\citenamefont {Rupp}\ \emph {et~al.}(2012)\citenamefont {Rupp},
  \citenamefont {Tkatchenko}, \citenamefont {M\"uller},\ and\ \citenamefont
  {von Lilienfeld}}]{PhysRevLett.108.058301}%
  \BibitemOpen
  \bibfield  {author} {\bibinfo {author} {\bibfnamefont {M.}~\bibnamefont
  {Rupp}}, \bibinfo {author} {\bibfnamefont {A.}~\bibnamefont {Tkatchenko}},
  \bibinfo {author} {\bibfnamefont {K.-R.}\ \bibnamefont {M\"uller}}, \ and\
  \bibinfo {author} {\bibfnamefont {O.~A.}\ \bibnamefont {von Lilienfeld}},\
  }\href {\doibase 10.1103/PhysRevLett.108.058301} {\bibfield  {journal}
  {\bibinfo  {journal} {Phys. Rev. Lett.}\ }\textbf {\bibinfo {volume} {108}},\
  \bibinfo {pages} {058301} (\bibinfo {year} {2012})}\BibitemShut {NoStop}%
\bibitem [{\citenamefont {Bartok}\ \emph {et~al.}(2017)\citenamefont {Bartok},
  \citenamefont {De}, \citenamefont {Poelking}, \citenamefont {Bernstein},
  \citenamefont {Kermode}, \citenamefont {Csanyi},\ and\ \citenamefont
  {Ceriotti}}]{bartok2017machine}%
  \BibitemOpen
  \bibfield  {author} {\bibinfo {author} {\bibfnamefont {A.~P.}\ \bibnamefont
  {Bartok}}, \bibinfo {author} {\bibfnamefont {S.}~\bibnamefont {De}}, \bibinfo
  {author} {\bibfnamefont {C.}~\bibnamefont {Poelking}}, \bibinfo {author}
  {\bibfnamefont {N.}~\bibnamefont {Bernstein}}, \bibinfo {author}
  {\bibfnamefont {J.}~\bibnamefont {Kermode}}, \bibinfo {author} {\bibfnamefont
  {G.}~\bibnamefont {Csanyi}}, \ and\ \bibinfo {author} {\bibfnamefont
  {M.}~\bibnamefont {Ceriotti}},\ }\href@noop {} {\bibfield  {journal}
  {\bibinfo  {journal} {arXiv preprint arXiv:1706.00179}\ } (\bibinfo {year}
  {2017})}\BibitemShut {NoStop}%
\bibitem [{\citenamefont {Gao}\ \emph {et~al.}(2016)\citenamefont {Gao},
  \citenamefont {Li}, \citenamefont {Li}, \citenamefont {Li}, \citenamefont
  {Fang}, \citenamefont {Li}, \citenamefont {Hu}, \citenamefont {Lu},\ and\
  \citenamefont {Su}}]{gao2016machine}%
  \BibitemOpen
  \bibfield  {author} {\bibinfo {author} {\bibfnamefont {T.}~\bibnamefont
  {Gao}}, \bibinfo {author} {\bibfnamefont {H.}~\bibnamefont {Li}}, \bibinfo
  {author} {\bibfnamefont {W.}~\bibnamefont {Li}}, \bibinfo {author}
  {\bibfnamefont {L.}~\bibnamefont {Li}}, \bibinfo {author} {\bibfnamefont
  {C.}~\bibnamefont {Fang}}, \bibinfo {author} {\bibfnamefont {H.}~\bibnamefont
  {Li}}, \bibinfo {author} {\bibfnamefont {L.}~\bibnamefont {Hu}}, \bibinfo
  {author} {\bibfnamefont {Y.}~\bibnamefont {Lu}}, \ and\ \bibinfo {author}
  {\bibfnamefont {Z.-M.}\ \bibnamefont {Su}},\ }\href@noop {} {\bibfield
  {journal} {\bibinfo  {journal} {Journal of cheminformatics}\ }\textbf
  {\bibinfo {volume} {8}},\ \bibinfo {pages} {24} (\bibinfo {year}
  {2016})}\BibitemShut {NoStop}%
\bibitem [{\citenamefont {Zhang}\ \emph {et~al.}(2018)\citenamefont {Zhang},
  \citenamefont {Shen},\ and\ \citenamefont {Zhai}}]{PhysRevLett.120.066401}%
  \BibitemOpen
  \bibfield  {author} {\bibinfo {author} {\bibfnamefont {P.}~\bibnamefont
  {Zhang}}, \bibinfo {author} {\bibfnamefont {H.}~\bibnamefont {Shen}}, \ and\
  \bibinfo {author} {\bibfnamefont {H.}~\bibnamefont {Zhai}},\ }\href {\doibase
  10.1103/PhysRevLett.120.066401} {\bibfield  {journal} {\bibinfo  {journal}
  {Phys. Rev. Lett.}\ }\textbf {\bibinfo {volume} {120}},\ \bibinfo {pages}
  {066401} (\bibinfo {year} {2018})}\BibitemShut {NoStop}%
\bibitem [{\citenamefont {Yoshioka}\ \emph {et~al.}(2017)\citenamefont
  {Yoshioka}, \citenamefont {Akagi},\ and\ \citenamefont
  {Katsura}}]{yoshioka2017learning}%
  \BibitemOpen
  \bibfield  {author} {\bibinfo {author} {\bibfnamefont {N.}~\bibnamefont
  {Yoshioka}}, \bibinfo {author} {\bibfnamefont {Y.}~\bibnamefont {Akagi}}, \
  and\ \bibinfo {author} {\bibfnamefont {H.}~\bibnamefont {Katsura}},\
  }\href@noop {} {\bibfield  {journal} {\bibinfo  {journal} {arXiv preprint
  arXiv:1709.05790}\ } (\bibinfo {year} {2017})}\BibitemShut {NoStop}%
\bibitem [{\citenamefont {Krizhevsky}\ \emph {et~al.}(2012)\citenamefont
  {Krizhevsky}, \citenamefont {Sutskever},\ and\ \citenamefont
  {Hinton}}]{alexnet2012}%
  \BibitemOpen
  \bibfield  {author} {\bibinfo {author} {\bibfnamefont {A.}~\bibnamefont
  {Krizhevsky}}, \bibinfo {author} {\bibfnamefont {I.}~\bibnamefont
  {Sutskever}}, \ and\ \bibinfo {author} {\bibfnamefont {G.~E.}\ \bibnamefont
  {Hinton}},\ }in\ \href
  {http://papers.nips.cc/paper/4824-imagenet-classification-with-deep-convolutional-neural-networks.pdf}
  {\emph {\bibinfo {booktitle} {Advances in Neural Information Processing
  Systems 25}}}\ (\bibinfo {year} {2012})\ pp.\ \bibinfo {pages}
  {1097--1105}\BibitemShut {NoStop}%
\bibitem [{\citenamefont {Dede}\ and\ \citenamefont
  {Sazlı}(2010)}]{Dede20107}%
  \BibitemOpen
  \bibfield  {author} {\bibinfo {author} {\bibfnamefont {G.}~\bibnamefont
  {Dede}}\ and\ \bibinfo {author} {\bibfnamefont {M.~H.}\ \bibnamefont
  {Sazlı}},\ }\href {\doibase https://doi.org/10.1016/j.dsp.2009.10.004}
  {\bibfield  {journal} {\bibinfo  {journal} {Digital Signal Processing}\
  }\textbf {\bibinfo {volume} {20}},\ \bibinfo {pages} {763 } (\bibinfo {year}
  {2010})}\BibitemShut {NoStop}%
\bibitem [{\citenamefont {Lecun}\ \emph {et~al.}(1998)\citenamefont {Lecun},
  \citenamefont {Bottou}, \citenamefont {Bengio},\ and\ \citenamefont
  {Haffner}}]{Lecun1998}%
  \BibitemOpen
  \bibfield  {author} {\bibinfo {author} {\bibfnamefont {Y.}~\bibnamefont
  {Lecun}}, \bibinfo {author} {\bibfnamefont {L.}~\bibnamefont {Bottou}},
  \bibinfo {author} {\bibfnamefont {Y.}~\bibnamefont {Bengio}}, \ and\ \bibinfo
  {author} {\bibfnamefont {P.}~\bibnamefont {Haffner}},\ }\href {\doibase
  10.1109/5.726791} {\bibfield  {journal} {\bibinfo  {journal} {Proc. IEEE}\
  }\textbf {\bibinfo {volume} {86}},\ \bibinfo {pages} {2278} (\bibinfo {year}
  {1998})}\BibitemShut {NoStop}%
\bibitem [{\citenamefont {Goldberg}(2015)}]{Goldberg2015}%
  \BibitemOpen
  \bibfield  {author} {\bibinfo {author} {\bibfnamefont {Y.}~\bibnamefont
  {Goldberg}},\ }\href {https://arxiv.org/pdf/1510.00726.pdf} {\bibfield
  {journal} {\bibinfo  {journal} {arXiv}\ } (\bibinfo {year} {2015})},\ \Eprint
  {http://arxiv.org/abs/arXiv:1510.00726v1} {arXiv:1510.00726v1} \BibitemShut
  {NoStop}%
\bibitem [{\citenamefont {Bengio}\ \emph {et~al.}(2003)\citenamefont {Bengio},
  \citenamefont {Ducharme}, \citenamefont {Vincent},\ and\ \citenamefont
  {Janvin}}]{Bengio2003}%
  \BibitemOpen
  \bibfield  {author} {\bibinfo {author} {\bibfnamefont {Y.}~\bibnamefont
  {Bengio}}, \bibinfo {author} {\bibfnamefont {R.}~\bibnamefont {Ducharme}},
  \bibinfo {author} {\bibfnamefont {P.}~\bibnamefont {Vincent}}, \ and\
  \bibinfo {author} {\bibfnamefont {C.}~\bibnamefont {Janvin}},\ }\href
  {http://dl.acm.org/citation.cfm?id=944919.944966} {\bibfield  {journal}
  {\bibinfo  {journal} {J. Mach. Learn. Res.}\ }\textbf {\bibinfo {volume}
  {3}},\ \bibinfo {pages} {1137} (\bibinfo {year} {2003})}\BibitemShut
  {NoStop}%
\bibitem [{\citenamefont {Schaul}\ \emph {et~al.}(2010)\citenamefont {Schaul},
  \citenamefont {Bayer}, \citenamefont {Wierstra}, \citenamefont {Sun},
  \citenamefont {Felder}, \citenamefont {Sehnke}, \citenamefont
  {R{\"u}ckstie{\ss}},\ and\ \citenamefont {Schmidhuber}}]{pybrain2010jmlr}%
  \BibitemOpen
  \bibfield  {author} {\bibinfo {author} {\bibfnamefont {T.}~\bibnamefont
  {Schaul}}, \bibinfo {author} {\bibfnamefont {J.}~\bibnamefont {Bayer}},
  \bibinfo {author} {\bibfnamefont {D.}~\bibnamefont {Wierstra}}, \bibinfo
  {author} {\bibfnamefont {Y.}~\bibnamefont {Sun}}, \bibinfo {author}
  {\bibfnamefont {M.}~\bibnamefont {Felder}}, \bibinfo {author} {\bibfnamefont
  {F.}~\bibnamefont {Sehnke}}, \bibinfo {author} {\bibfnamefont
  {T.}~\bibnamefont {R{\"u}ckstie{\ss}}}, \ and\ \bibinfo {author}
  {\bibfnamefont {J.}~\bibnamefont {Schmidhuber}},\ }\href@noop {} {\bibfield
  {journal} {\bibinfo  {journal} {Journal of Machine Learning Research}\
  }\textbf {\bibinfo {volume} {11}},\ \bibinfo {pages} {743} (\bibinfo {year}
  {2010})}\BibitemShut {NoStop}%
\bibitem [{\citenamefont {Solomonoff}(1957)}]{Solomonoff1957}%
  \BibitemOpen
  \bibfield  {author} {\bibinfo {author} {\bibfnamefont {R.}~\bibnamefont
  {Solomonoff}},\ }\href {http://raysolomonoff.com/publications/An Inductive
  Inference Machine1957.pdf} {\bibfield  {journal} {\bibinfo  {journal} {IRE
  Convention Record}\ }\textbf {\bibinfo {volume} {2}},\ \bibinfo {pages} {56}
  (\bibinfo {year} {1957})}\BibitemShut {NoStop}%
\bibitem [{\citenamefont {Samuel}(1959)}]{Samuel1959}%
  \BibitemOpen
  \bibfield  {author} {\bibinfo {author} {\bibfnamefont {A.~L.}\ \bibnamefont
  {Samuel}},\ }\href {\doibase 10.1147/rd.441.0206} {\bibfield  {journal}
  {\bibinfo  {journal} {IBM Journal of Research and Development}\ }\textbf
  {\bibinfo {volume} {3}},\ \bibinfo {pages} {206} (\bibinfo {year}
  {1959})}\BibitemShut {NoStop}%
\bibitem [{\citenamefont {Rosenblatt}(1958)}]{Rosenblatt1958}%
  \BibitemOpen
  \bibfield  {author} {\bibinfo {author} {\bibfnamefont {F.}~\bibnamefont
  {Rosenblatt}},\ }\href {\doibase 10.1037/h0042519} {\bibfield  {journal}
  {\bibinfo  {journal} {Psychological Review}\ }\textbf {\bibinfo {volume}
  {65}},\ \bibinfo {pages} {386} (\bibinfo {year} {1958})}\BibitemShut
  {NoStop}%
\bibitem [{\citenamefont {Hodgkin}\ and\ \citenamefont
  {Huxley}(1952)}]{Hodgkin1952}%
  \BibitemOpen
  \bibfield  {author} {\bibinfo {author} {\bibfnamefont {A.~L.}\ \bibnamefont
  {Hodgkin}}\ and\ \bibinfo {author} {\bibfnamefont {A.~F.}\ \bibnamefont
  {Huxley}},\ }\href@noop {} {\bibfield  {journal} {\bibinfo  {journal} {J.
  Physiol.}\ }\textbf {\bibinfo {volume} {117}},\ \bibinfo {pages} {500}
  (\bibinfo {year} {1952})}\BibitemShut {NoStop}%
\bibitem [{\citenamefont {Hopfield}(1982)}]{Hopfield1982}%
  \BibitemOpen
  \bibfield  {author} {\bibinfo {author} {\bibfnamefont {J.~J.}\ \bibnamefont
  {Hopfield}},\ }\href@noop {} {\bibfield  {journal} {\bibinfo  {journal}
  {PNAS}\ }\textbf {\bibinfo {volume} {79}},\ \bibinfo {pages} {2554} (\bibinfo
  {year} {1982})}\BibitemShut {NoStop}%
\bibitem [{\citenamefont {Rumelhart}\ \emph {et~al.}(1986)\citenamefont
  {Rumelhart}, \citenamefont {Hinton},\ and\ \citenamefont
  {Williams}}]{Rumelhart1986}%
  \BibitemOpen
  \bibfield  {author} {\bibinfo {author} {\bibfnamefont {D.~E.}\ \bibnamefont
  {Rumelhart}}, \bibinfo {author} {\bibfnamefont {G.~E.}\ \bibnamefont
  {Hinton}}, \ and\ \bibinfo {author} {\bibfnamefont {R.~J.}\ \bibnamefont
  {Williams}},\ }\href {\doibase 10.1038/323533a0} {\bibfield  {journal}
  {\bibinfo  {journal} {Nature}\ }\textbf {\bibinfo {volume} {323}},\ \bibinfo
  {pages} {533} (\bibinfo {year} {1986})}\BibitemShut {NoStop}%
\bibitem [{\citenamefont {Wei\ss{}e}\ \emph {et~al.}(2006)\citenamefont
  {Wei\ss{}e}, \citenamefont {Wellein}, \citenamefont {Alvermann},\ and\
  \citenamefont {Fehske}}]{RevModPhys.78.275}%
  \BibitemOpen
  \bibfield  {author} {\bibinfo {author} {\bibfnamefont {A.}~\bibnamefont
  {Wei\ss{}e}}, \bibinfo {author} {\bibfnamefont {G.}~\bibnamefont {Wellein}},
  \bibinfo {author} {\bibfnamefont {A.}~\bibnamefont {Alvermann}}, \ and\
  \bibinfo {author} {\bibfnamefont {H.}~\bibnamefont {Fehske}},\ }\href
  {\doibase 10.1103/RevModPhys.78.275} {\bibfield  {journal} {\bibinfo
  {journal} {Rev. Mod. Phys.}\ }\textbf {\bibinfo {volume} {78}},\ \bibinfo
  {pages} {275} (\bibinfo {year} {2006})}\BibitemShut {NoStop}%
\bibitem [{\citenamefont {Budich}\ and\ \citenamefont
  {Ardonne}(2013)}]{PhysRevB.88.075419}%
  \BibitemOpen
  \bibfield  {author} {\bibinfo {author} {\bibfnamefont {J.~C.}\ \bibnamefont
  {Budich}}\ and\ \bibinfo {author} {\bibfnamefont {E.}~\bibnamefont
  {Ardonne}},\ }\href {\doibase 10.1103/PhysRevB.88.075419} {\bibfield
  {journal} {\bibinfo  {journal} {Phys. Rev. B}\ }\textbf {\bibinfo {volume}
  {88}},\ \bibinfo {pages} {075419} (\bibinfo {year} {2013})}\BibitemShut
  {NoStop}%
\bibitem [{\citenamefont {Lutchyn}\ \emph {et~al.}(2010)\citenamefont
  {Lutchyn}, \citenamefont {Sau},\ and\ \citenamefont
  {Das~Sarma}}]{PhysRevLett.105.077001}%
  \BibitemOpen
  \bibfield  {author} {\bibinfo {author} {\bibfnamefont {R.~M.}\ \bibnamefont
  {Lutchyn}}, \bibinfo {author} {\bibfnamefont {J.~D.}\ \bibnamefont {Sau}}, \
  and\ \bibinfo {author} {\bibfnamefont {S.}~\bibnamefont {Das~Sarma}},\ }\href
  {\doibase 10.1103/PhysRevLett.105.077001} {\bibfield  {journal} {\bibinfo
  {journal} {Phys. Rev. Lett.}\ }\textbf {\bibinfo {volume} {105}},\ \bibinfo
  {pages} {077001} (\bibinfo {year} {2010})}\BibitemShut {NoStop}%
\bibitem [{\citenamefont {Oreg}\ \emph {et~al.}(2010)\citenamefont {Oreg},
  \citenamefont {Refael},\ and\ \citenamefont {von
  Oppen}}]{PhysRevLett.105.177002}%
  \BibitemOpen
  \bibfield  {author} {\bibinfo {author} {\bibfnamefont {Y.}~\bibnamefont
  {Oreg}}, \bibinfo {author} {\bibfnamefont {G.}~\bibnamefont {Refael}}, \ and\
  \bibinfo {author} {\bibfnamefont {F.}~\bibnamefont {von Oppen}},\ }\href
  {\doibase 10.1103/PhysRevLett.105.177002} {\bibfield  {journal} {\bibinfo
  {journal} {Phys. Rev. Lett.}\ }\textbf {\bibinfo {volume} {105}},\ \bibinfo
  {pages} {177002} (\bibinfo {year} {2010})}\BibitemShut {NoStop}%
\bibitem [{\citenamefont {Mourik}\ \emph {et~al.}(2012)\citenamefont {Mourik},
  \citenamefont {Zuo}, \citenamefont {Frolov}, \citenamefont {Plissard},
  \citenamefont {Bakkers},\ and\ \citenamefont
  {Kouwenhoven}}]{mourik2012signatures}%
  \BibitemOpen
  \bibfield  {author} {\bibinfo {author} {\bibfnamefont {V.}~\bibnamefont
  {Mourik}}, \bibinfo {author} {\bibfnamefont {K.}~\bibnamefont {Zuo}},
  \bibinfo {author} {\bibfnamefont {S.~M.}\ \bibnamefont {Frolov}}, \bibinfo
  {author} {\bibfnamefont {S.}~\bibnamefont {Plissard}}, \bibinfo {author}
  {\bibfnamefont {E.}~\bibnamefont {Bakkers}}, \ and\ \bibinfo {author}
  {\bibfnamefont {L.~P.}\ \bibnamefont {Kouwenhoven}},\ }\href@noop {}
  {\bibfield  {journal} {\bibinfo  {journal} {Science}\ }\textbf {\bibinfo
  {volume} {336}},\ \bibinfo {pages} {1003} (\bibinfo {year}
  {2012})}\BibitemShut {NoStop}%
\bibitem [{\citenamefont {Stanescu}\ \emph {et~al.}(2011)\citenamefont
  {Stanescu}, \citenamefont {Lutchyn},\ and\ \citenamefont
  {Das~Sarma}}]{PhysRevB.84.144522}%
  \BibitemOpen
  \bibfield  {author} {\bibinfo {author} {\bibfnamefont {T.~D.}\ \bibnamefont
  {Stanescu}}, \bibinfo {author} {\bibfnamefont {R.~M.}\ \bibnamefont
  {Lutchyn}}, \ and\ \bibinfo {author} {\bibfnamefont {S.}~\bibnamefont
  {Das~Sarma}},\ }\href {\doibase 10.1103/PhysRevB.84.144522} {\bibfield
  {journal} {\bibinfo  {journal} {Phys. Rev. B}\ }\textbf {\bibinfo {volume}
  {84}},\ \bibinfo {pages} {144522} (\bibinfo {year} {2011})}\BibitemShut
  {NoStop}%
\bibitem [{\citenamefont {Lutchyn}\ \emph {et~al.}(2011)\citenamefont
  {Lutchyn}, \citenamefont {Stanescu},\ and\ \citenamefont
  {Das~Sarma}}]{PhysRevLett.106.127001}%
  \BibitemOpen
  \bibfield  {author} {\bibinfo {author} {\bibfnamefont {R.~M.}\ \bibnamefont
  {Lutchyn}}, \bibinfo {author} {\bibfnamefont {T.~D.}\ \bibnamefont
  {Stanescu}}, \ and\ \bibinfo {author} {\bibfnamefont {S.}~\bibnamefont
  {Das~Sarma}},\ }\href {\doibase 10.1103/PhysRevLett.106.127001} {\bibfield
  {journal} {\bibinfo  {journal} {Phys. Rev. Lett.}\ }\textbf {\bibinfo
  {volume} {106}},\ \bibinfo {pages} {127001} (\bibinfo {year}
  {2011})}\BibitemShut {NoStop}%
\bibitem [{\citenamefont {Lutchyn}\ \emph {et~al.}(2017)\citenamefont
  {Lutchyn}, \citenamefont {Bakkers}, \citenamefont {Kouwenhoven},
  \citenamefont {Krogstrup}, \citenamefont {Marcus},\ and\ \citenamefont
  {Oreg}}]{lutchyn2017realizing}%
  \BibitemOpen
  \bibfield  {author} {\bibinfo {author} {\bibfnamefont {R.}~\bibnamefont
  {Lutchyn}}, \bibinfo {author} {\bibfnamefont {E.}~\bibnamefont {Bakkers}},
  \bibinfo {author} {\bibfnamefont {L.}~\bibnamefont {Kouwenhoven}}, \bibinfo
  {author} {\bibfnamefont {P.}~\bibnamefont {Krogstrup}}, \bibinfo {author}
  {\bibfnamefont {C.}~\bibnamefont {Marcus}}, \ and\ \bibinfo {author}
  {\bibfnamefont {Y.}~\bibnamefont {Oreg}},\ }\href@noop {} {\bibfield
  {journal} {\bibinfo  {journal} {arXiv preprint arXiv:1707.04899}\ } (\bibinfo
  {year} {2017})}\BibitemShut {NoStop}%
\bibitem [{\citenamefont {Aguado}(2017)}]{aguado2017majorana}%
  \BibitemOpen
  \bibfield  {author} {\bibinfo {author} {\bibfnamefont {R.}~\bibnamefont
  {Aguado}},\ }\href@noop {} {\bibfield  {journal} {\bibinfo  {journal} {arXiv
  preprint arXiv:1711.00011}\ } (\bibinfo {year} {2017})}\BibitemShut {NoStop}%
\bibitem [{\citenamefont {Qiao}\ \emph {et~al.}(2010)\citenamefont {Qiao},
  \citenamefont {Yang}, \citenamefont {Feng}, \citenamefont {Tse},
  \citenamefont {Ding}, \citenamefont {Yao}, \citenamefont {Wang},\ and\
  \citenamefont {Niu}}]{Qiao2010}%
  \BibitemOpen
  \bibfield  {author} {\bibinfo {author} {\bibfnamefont {Z.}~\bibnamefont
  {Qiao}}, \bibinfo {author} {\bibfnamefont {S.~A.}\ \bibnamefont {Yang}},
  \bibinfo {author} {\bibfnamefont {W.}~\bibnamefont {Feng}}, \bibinfo {author}
  {\bibfnamefont {W.-K.}\ \bibnamefont {Tse}}, \bibinfo {author} {\bibfnamefont
  {J.}~\bibnamefont {Ding}}, \bibinfo {author} {\bibfnamefont {Y.}~\bibnamefont
  {Yao}}, \bibinfo {author} {\bibfnamefont {J.}~\bibnamefont {Wang}}, \ and\
  \bibinfo {author} {\bibfnamefont {Q.}~\bibnamefont {Niu}},\ }\href {\doibase
  10.1103/PhysRevB.82.161414} {\bibfield  {journal} {\bibinfo  {journal} {Phys.
  Rev. B}\ }\textbf {\bibinfo {volume} {82}},\ \bibinfo {pages} {161414}
  (\bibinfo {year} {2010})},\ \Eprint {http://arxiv.org/abs/arXiv:1005.1672}
  {arXiv:arXiv:1005.1672} \BibitemShut {NoStop}%
\bibitem [{\citenamefont {Min}\ \emph {et~al.}(2006)\citenamefont {Min},
  \citenamefont {Hill}, \citenamefont {Sinitsyn}, \citenamefont {Sahu},
  \citenamefont {Kleinman},\ and\ \citenamefont {MacDonald}}]{Min2006}%
  \BibitemOpen
  \bibfield  {author} {\bibinfo {author} {\bibfnamefont {H.}~\bibnamefont
  {Min}}, \bibinfo {author} {\bibfnamefont {J.~E.}\ \bibnamefont {Hill}},
  \bibinfo {author} {\bibfnamefont {N.~A.}\ \bibnamefont {Sinitsyn}}, \bibinfo
  {author} {\bibfnamefont {B.~R.}\ \bibnamefont {Sahu}}, \bibinfo {author}
  {\bibfnamefont {L.}~\bibnamefont {Kleinman}}, \ and\ \bibinfo {author}
  {\bibfnamefont {A.~H.}\ \bibnamefont {MacDonald}},\ }\href {\doibase
  10.1103/PhysRevB.74.165310} {\bibfield  {journal} {\bibinfo  {journal}
  {Physical Review B}\ }\textbf {\bibinfo {volume} {74}},\ \bibinfo {pages}
  {165310} (\bibinfo {year} {2006})},\ \Eprint {http://arxiv.org/abs/0606504}
  {arXiv:0606504} \BibitemShut {NoStop}%
\bibitem [{\citenamefont {Lado}\ \emph {et~al.}(2013)\citenamefont {Lado},
  \citenamefont {Gonz\'alez},\ and\ \citenamefont
  {Fern\'andez-Rossier}}]{PhysRevB.88.035448}%
  \BibitemOpen
  \bibfield  {author} {\bibinfo {author} {\bibfnamefont {J.~L.}\ \bibnamefont
  {Lado}}, \bibinfo {author} {\bibfnamefont {J.~W.}\ \bibnamefont
  {Gonz\'alez}}, \ and\ \bibinfo {author} {\bibfnamefont {J.}~\bibnamefont
  {Fern\'andez-Rossier}},\ }\href {\doibase 10.1103/PhysRevB.88.035448}
  {\bibfield  {journal} {\bibinfo  {journal} {Phys. Rev. B}\ }\textbf {\bibinfo
  {volume} {88}},\ \bibinfo {pages} {035448} (\bibinfo {year}
  {2013})}\BibitemShut {NoStop}%
\bibitem [{\citenamefont {Tang}\ \emph {et~al.}(2017)\citenamefont {Tang},
  \citenamefont {Cheng}, \citenamefont {Aldosary}, \citenamefont {Wang},
  \citenamefont {Jiang}, \citenamefont {Watanabe}, \citenamefont {Taniguchi},
  \citenamefont {Bockrath},\ and\ \citenamefont {Shi}}]{tang2017approaching}%
  \BibitemOpen
  \bibfield  {author} {\bibinfo {author} {\bibfnamefont {C.}~\bibnamefont
  {Tang}}, \bibinfo {author} {\bibfnamefont {B.}~\bibnamefont {Cheng}},
  \bibinfo {author} {\bibfnamefont {M.}~\bibnamefont {Aldosary}}, \bibinfo
  {author} {\bibfnamefont {Z.}~\bibnamefont {Wang}}, \bibinfo {author}
  {\bibfnamefont {Z.}~\bibnamefont {Jiang}}, \bibinfo {author} {\bibfnamefont
  {K.}~\bibnamefont {Watanabe}}, \bibinfo {author} {\bibfnamefont
  {T.}~\bibnamefont {Taniguchi}}, \bibinfo {author} {\bibfnamefont
  {M.}~\bibnamefont {Bockrath}}, \ and\ \bibinfo {author} {\bibfnamefont
  {J.}~\bibnamefont {Shi}},\ }\href@noop {} {\bibfield  {journal} {\bibinfo
  {journal} {arXiv preprint arXiv:1710.04179}\ } (\bibinfo {year}
  {2017})}\BibitemShut {NoStop}%
\bibitem [{\citenamefont {Wang}\ \emph {et~al.}(2015)\citenamefont {Wang},
  \citenamefont {Tang}, \citenamefont {Sachs}, \citenamefont {Barlas},\ and\
  \citenamefont {Shi}}]{PhysRevLett.114.016603}%
  \BibitemOpen
  \bibfield  {author} {\bibinfo {author} {\bibfnamefont {Z.}~\bibnamefont
  {Wang}}, \bibinfo {author} {\bibfnamefont {C.}~\bibnamefont {Tang}}, \bibinfo
  {author} {\bibfnamefont {R.}~\bibnamefont {Sachs}}, \bibinfo {author}
  {\bibfnamefont {Y.}~\bibnamefont {Barlas}}, \ and\ \bibinfo {author}
  {\bibfnamefont {J.}~\bibnamefont {Shi}},\ }\href {\doibase
  10.1103/PhysRevLett.114.016603} {\bibfield  {journal} {\bibinfo  {journal}
  {Phys. Rev. Lett.}\ }\textbf {\bibinfo {volume} {114}},\ \bibinfo {pages}
  {016603} (\bibinfo {year} {2015})}\BibitemShut {NoStop}%
\bibitem [{\citenamefont {Su}\ \emph {et~al.}(2017)\citenamefont {Su},
  \citenamefont {Barlas}, \citenamefont {Li}, \citenamefont {Shi},\ and\
  \citenamefont {Lake}}]{PhysRevB.95.075418}%
  \BibitemOpen
  \bibfield  {author} {\bibinfo {author} {\bibfnamefont {S.}~\bibnamefont
  {Su}}, \bibinfo {author} {\bibfnamefont {Y.}~\bibnamefont {Barlas}}, \bibinfo
  {author} {\bibfnamefont {J.}~\bibnamefont {Li}}, \bibinfo {author}
  {\bibfnamefont {J.}~\bibnamefont {Shi}}, \ and\ \bibinfo {author}
  {\bibfnamefont {R.~K.}\ \bibnamefont {Lake}},\ }\href {\doibase
  10.1103/PhysRevB.95.075418} {\bibfield  {journal} {\bibinfo  {journal} {Phys.
  Rev. B}\ }\textbf {\bibinfo {volume} {95}},\ \bibinfo {pages} {075418}
  (\bibinfo {year} {2017})}\BibitemShut {NoStop}%
\bibitem [{\citenamefont {Huang}\ \emph {et~al.}(2017)\citenamefont {Huang},
  \citenamefont {Clark}, \citenamefont {Navarro-Moratalla}, \citenamefont
  {Klein}, \citenamefont {Cheng}, \citenamefont {Seyler}, \citenamefont
  {Zhong}, \citenamefont {Schmidgall}, \citenamefont {McGuire}, \citenamefont
  {Cobden} \emph {et~al.}}]{huang2017layer}%
  \BibitemOpen
  \bibfield  {author} {\bibinfo {author} {\bibfnamefont {B.}~\bibnamefont
  {Huang}}, \bibinfo {author} {\bibfnamefont {G.}~\bibnamefont {Clark}},
  \bibinfo {author} {\bibfnamefont {E.}~\bibnamefont {Navarro-Moratalla}},
  \bibinfo {author} {\bibfnamefont {D.~R.}\ \bibnamefont {Klein}}, \bibinfo
  {author} {\bibfnamefont {R.}~\bibnamefont {Cheng}}, \bibinfo {author}
  {\bibfnamefont {K.~L.}\ \bibnamefont {Seyler}}, \bibinfo {author}
  {\bibfnamefont {D.}~\bibnamefont {Zhong}}, \bibinfo {author} {\bibfnamefont
  {E.}~\bibnamefont {Schmidgall}}, \bibinfo {author} {\bibfnamefont {M.~A.}\
  \bibnamefont {McGuire}}, \bibinfo {author} {\bibfnamefont {D.~H.}\
  \bibnamefont {Cobden}},  \emph {et~al.},\ }\href@noop {} {\bibfield
  {journal} {\bibinfo  {journal} {Nature}\ }\textbf {\bibinfo {volume} {546}},\
  \bibinfo {pages} {270} (\bibinfo {year} {2017})}\BibitemShut {NoStop}%
\bibitem [{\citenamefont {Zhang}\ \emph {et~al.}(2017)\citenamefont {Zhang},
  \citenamefont {Zhao}, \citenamefont {Zhou}, \citenamefont {Xue},
  \citenamefont {Ma},\ and\ \citenamefont {Yang}}]{zhang2017strong}%
  \BibitemOpen
  \bibfield  {author} {\bibinfo {author} {\bibfnamefont {J.}~\bibnamefont
  {Zhang}}, \bibinfo {author} {\bibfnamefont {B.}~\bibnamefont {Zhao}},
  \bibinfo {author} {\bibfnamefont {T.}~\bibnamefont {Zhou}}, \bibinfo {author}
  {\bibfnamefont {Y.}~\bibnamefont {Xue}}, \bibinfo {author} {\bibfnamefont
  {C.}~\bibnamefont {Ma}}, \ and\ \bibinfo {author} {\bibfnamefont
  {Z.}~\bibnamefont {Yang}},\ }\href@noop {} {\bibfield  {journal} {\bibinfo
  {journal} {arXiv preprint arXiv:1710.06324}\ } (\bibinfo {year}
  {2017})}\BibitemShut {NoStop}%
\end{thebibliography}%

\end{document}